# Ammoniated electron as a solvent stabilized multimer radical anion.

Ilya A. Shkrob

*Chemistry Division , Argonne National Laboratory, 9700 S. Cass Ave,*

*Argonne, IL 60439*

**Abstract**

The excess electron in liquid ammonia ("ammoniated electron") is commonly viewed as a cavity electron in which the *s*-type wave function fills the interstitial void between 6-9 ammonia molecules. Here we examine an alternative model in which the ammoniated electron is regarded as a solvent stabilized multimer radical anion, as was originally suggested by Symons [*Chem. Soc. Rev.* **1976**, *5*, 337]. In this model, most of the excess electron density resides in the frontier orbitals of *N* atoms in the ammonia molecules forming the solvation cavity; a fraction of this spin density is transferred to the molecules in the second solvation shell. The cavity is formed due to the repulsion between negatively charged solvent molecules. Using density functional theory calculations for small ammonia cluster anions in the gas phase, it is demonstrated that such core anions would semi-quantitatively account for the observed pattern of Knight shifts for $^1$H and $^{14}$N nuclei observed by NMR spectroscopy and the downshifted stretching and bending modes observed by infrared spectroscopy. It is speculated that the excess electrons in other aprotic solvents (but not in water and alcohols) might be, in this respect, analogous to the ammoniated electron, with substantial transfer of the spin density into the frontier *N* and *C* orbitals of methyl, amino, and amide groups forming the solvation cavity.

___________________________________________________________

[a] Author to whom correspondence should be addressed; electronic mail: shkrob@anl.gov.

1. Introduction.

While the solvated electron in liquid ammonia (also known as "ammoniated electron", $e_{am}^-$) is the first known example [1] of a stable excess electron in any liquid, [2,3] complete understanding of its structure and properties remains elusive. Most of the theories of electron solvation are one-electron models in which a single quantum mechanical particle, the excess electron, interacts with the solvent molecules (that are treated classically) by means of an effective potential. This idealized particle-in-a-box approach has been the standard fixture of all successful theories for electron solvation, from the original (static) dielectric continuum [4,5,6] and semicontinuum models [7-11] to the latest advanced models [12-16] in which the solvent dynamics are explicitly treated. For the ammoniated electron, the one-electron models were first suggested by Ogg, [4] and further developed by Jortner [5] and Kestner. [6] More recent examples of this approach are Feynman's Path Integral Monte-Carlo (PIMC) calculations of Klein and co-workers [17-20] and Rodriguez et al. [21] and Quantum Path Integral Molecular Dynamics (QUPID) calculations for large anion clusters by Barnett et al. [22] (the dielectric continuum models were adapted for such clusters in refs. 23).

These theories suggest that the ammoniated electron is localized in a cavity comprised of 8-9 ammonia molecules with the gyration radius variously estimated between 2 and 4 Å. As suggested by PIMC calculations of Sprik, Impey and Klein, [17] the pair correlation function for center-of-mass (c) electron-H distribution has a peak at 2.1 Å and the c-N distribution has a peak at 2.9 Å. These estimates depend on the choice of (unknown) pseudopotential; e.g., Marchi et al. [20] and Rodriguez et al. [21] give for these two peaks 3.4 and 4 Å, respectively. The volume change on electron solvation at 1 atm is ca. 100 cm$^3$/mol which is three times the volume occupied by a solvent molecule. [20] This volume corresponds to a sphere of radius of 3.4 Å. Using dielectric continuum models, Jortner [5] estimated the hard core radius of the spherical cavity as 3 Å, while Kestner [6] obtained a lower estimate of 1.7-2.2 Å. The cavity is formed by dangling N-H groups, and the $s$-like wavefunction of the ground state excess electron is fully contained within the cavity (at 1 atm). In the one-electron models, the cavity is formed due to the Pauli exclusion of the excess electron by the valence electrons in ammonia molecules (which have no electron affinity in the gas phase). Some model calculations suggested that there is a preferential orientation of one of the N-H bonds towards the center of the cavity, whereas other calculations indicated that an orientation in which all three N-H bonds of the molecule are turned towards the cavity center were equally likely (see Abramczyk and Kroh [8] for more detail). The differences in the results from the different models arise from the relative weakness of N..H-N hydrogen bonding in ammonia, as compared to water. [24]

The strongest support for one-electron models for $e_{am}^-$ is provided by the ease with which such models [e.g., 5-8,11,17,20,21] account for its broad absorption ($s \to p$) band in the near infrared (IR) that is centered at 0.8 eV and has an oscillator strength of 0.77. [6] Similar models were suggested for the (hydrated) solvated electron in liquid water and alcohols. Beginning in the early 1980s, most of the theoretical studies on electron solvation were carried out for the *hydrated* electron; [12-16] the interest in $e_{am}^-$ was sporadic, except for a brief period of time following the discovery of $am_n^-$ cluster anions in the gas phase, [22,23] by Haberland and co-workers [25] and others. [26] For water, there are many indications, both direct and indirect, that the one-electron picture does capture the essential physics, [27] at least to a first approximation i. e., the transfer of spin density to the frontier orbitals of oxygen is relatively small. For $e_{am}^-$, the bulk of the experimental evidence points to the contrary. Below, we briefly review this evidence.

Ammoniated electrons can be prepared by dissolving alkali metal in liquid ammonia. [2,3] When the concentration of the metal is lower than a few millimoles per dm$^3$, the properties of the excess electron do not depend on the type and the concentration of the metal, which suggests that alkali cations are not included in the cavity. [2,3,28] Due to the stability of the excess electron in ammonia, spectral data for $e_{am}^-$ can be obtained that are lacking for the excess electrons in hydroxylated, liquid solvents, such as water, where the lifetime of the electron is limited. [27] In particular, it is possible to determine the Knight shifts for $^1$H and $^{14}$N nuclei in the molecules that "solvate" $e_{am}^-$. [28,29,30] To our knowledge, such data exist for only two other solvents, methylamine [31] and hexamethyl phosphoramide (HMPA); [32] solutions of Na in these liquids [32-35] also yield stable solvated electrons that absorb in the IR [33,34] and exhibit a characteristic motional narrowed line in their Electron Paramagnetic Resonance (EPR) spectra. [33,35] The Knight shift $K_X$ of nuclear magnetic resonance (NMR) lines is due to the contact Fermi (isotropic) hyperfine interaction of the excess electron with the magnetic nuclei ($X$) in the solvent molecules; [2,3,28] it is the measure of spin density $|\phi_s(0)|^2_X$ in the $s$-type atomic orbitals (AOs) centered on a given nucleus $X$:

$$K_X = \frac{8\pi}{3N_e} \chi^P \Sigma_X |\phi_s(0)|^2_X \qquad (1)$$

where $N_e$ is the number density of ammoniated electrons and $\chi^P \approx \gamma_e^2/4k_BT$ is the (experimentally determined) [28] electron paramagnetic susceptibility, where $\gamma_e$ is the gyromagnetic ratio of the electron and $k_BT$ is the thermal energy. This shift can be converted into the sum

$$\Sigma_X A = \frac{8\pi}{3} \gamma_e \gamma_X \Sigma_X |\phi_s(0)|^2_X \qquad (2)$$

of isotropic hyperfine coupling constants (hfcc's) for all nuclei of type $X$ with gyromagnetic ratio $\gamma_X$ that interact with the excess electron. In ammonia, this calculation gives +110 G for $^{14}$N and -5.7 G for $^1$H (1 G = $10^{-4}$ T in field units is equivalent to 2.8 MHz in frequency units) for $\Sigma_X|\phi_s(0)|^2_X$ of +0.954 $a_0^{-3}$ and -3.56x10$^{-3}$ $a_0^{-3}$, [28] respectively, where $a_0 \approx 0.53$ Å is the Bohr radius. [3,29,31] The negative sign of the isotropic hfcc for protons was demonstrated by Lambert [36] using dynamic nuclear polarization experiments and then confirmed by direct NMR measurements. [29] Given that the atomic hfcc for the electron in the 2s orbital of $^{14}$N is +550 G, Symons [3] estimated that ≈20% of the total spin density of the excess electron is transferred to these $N$ 2s orbitals. This immediately suggests that $e^-_{am}$ is, in fact, a solvent stabilized radical anion in which the unpaired electron is shared by ammonia molecules; only a fraction of the total electron density resides in the interstitial cavity. Symon's estimate does not include the spin density transferred to $N$ 2p orbitals, as only $s$-type AOs contribute to the *isotropic* hfcc on $^{14}$N.

However, such a transfer can be inferred from the negative sign for proton hfcc's. [3] This sign presents a formidable challenge to one-electron models, as only positively valued hfcc's can be obtained in the absence of bond spin polarization. [37] Symons [3] suggested that the proton hfcc is negative due to the conjugation of *H 1s* and *N 2p* orbitals via a spin bond polarization mechanism involving the hybrid *N sp²* orbital, similar to that occurring in *p*-radicals. This polarization results in a negative contribution to the proton hfcc that cancels the positive contribution from the interaction with the cavity-filling electron density. For this mechanism to operate at all, there must be a substantial occupancy of the *N 2p* orbitals.

Additional evidence for the crucial involvement of nitrogen AOs comes from the EPR and electron spin echo spectroscopies and NMR relaxation studies of dilute sodium solutions reviewed in refs. 2, 3, 28, and 36. The studies of electron spin echo relaxation [38] and EPR linewidth [39] indicate that while the exchange of $^1$H for $^2$H does not decrease the electron relaxation times, the exchange of $^{14}$N for $^{15}$N decreases the $T_2$ time by 20-26%. [38] The relaxation times for $^1$H and $^2$H nuclei (for shifted NMR lines) are almost the same as in the bulk liquid, whereas the relaxation time for $^{14}$N is drastically shortened. [30,40,41,42] These observations suggest that the electron spin is strongly coupled to $^{14}$N nuclei; the contribution from the protons, even an anisotropic one (via dipole dipole coupling), is minor. Such a result appears to be inconsistent with a one-electron model, since the nitrogens are located *further* from the center of the solvation cavity than the protons, regardless of which orientation (bond or dipole) is preferred. Detailed models of $^1$H [36,41,42] and $^{14}$N [28,30] relaxation near $e^-_{am}$ that were developed in the late 1960s, following the original approach by Kaplan and Kittel, [40] suggested that the spin density is divided between 20-40 magnetically equivalent nitrogens (in the most advanced of these models suggested by Catterall, [28] there are only 3-13 such nitrogens). The uncertainty in these estimates is due to the unknown correlation time for the motion of solvent molecules near the cavity; it is this motion that causes the electron and nuclear spin relaxation.

Following the original suggestions of O'Reilly [9,30] and Land and O'Reilly, [10] these magnetic resonance data were initially construed to indicate that the electron in liquid ammonia is trapped inside a large bubble (similar to the electron bubble in liquid helium);

[43] the 20-40 ammonia molecules at the wall of this bubble were thought to share the spin density equally. Such a model was completely incompatible with optical spectroscopy and thermodynamics data and it was quickly abandoned in favor of a tight solvation cavity model advocated by Jortner [5] and Kestner. [6] In 1976, Symons [3] realized that the spin density does not have to be divided equally between the nitrogens. He speculated that $e^-_{am}$ does have the tight structure suggested by the one-electron models: the molecules are clustered around a small void (of ca. 4-6 Å in diameter) that is partially filled by the electron wavefunction. However, a fraction of the unpaired electron density is divided between 6 ammonia molecules in the first coordination shell (with an isotropic hfcc of ca. +12 G) and 12 molecules in the second shell (with a small hfcc of ca. +3 G). In 1979, Smith, Symons, and Wardman [44] used EPR to determine isotropic hfcc's for $^{14}N$ nuclei in ammoniated F-center on the surface of MgO; these hfcc's (ca. 11±0.5 G) were close to the predicted values [3] for $^{14}N$ nuclei in the first solvated shell of $e^-_{am}$. Further support of Symon's hypothesis was provided by *ab initio* calculations of dimer, trimer, and tetramer anion clusters by Newton [45] and Clark and Illing. [46] These calculations indicated large spin densities on *N* atoms and yielded small negative hfcc's for the protons via bond spin polarization, in a fashion predicted by Symons [6] (section 2).

In the early 1980s, the accuracy of Knight shift measurements for $^1H$ and $^{13}C$ nuclei was improved [29,31] and the temperature range of these measurements was increased. [29] These more recent measurements only strengthened the conclusions reached by the researchers in the 1960s and 1970s. In particular, Niibe and Nakamura [29] narrowed the estimate for the average coordination number of $e^-_{am}$ to ≈7 (assuming the magnetic equivalency of nitrogens) and re-estimated $\Sigma_H A$ to be ca. -11.7 G. Furthermore, Symons' approach [3] was successfully used to account for the observed Knight shifts on $^1H$, $^{14}N$, and $^{13}C$ nuclei for solvated electron in methylamine, [31] where both the amine and the methyl groups are at the cavity wall.

Apart from these magnetic resonance data, there is another structural aspect of the problem that has not yet been addressed within the framework of one-electron models. Raman spectra of the hydrated electron, recently obtained by Mathies and Tauber [47a,b] and Tahara and co-workers, [48] indicate a large downshift in the frequency of the O-H stretch (ca. 300 cm$^{-1}$) [47,48] and a smaller downshift for the H-O-H bend (ca. 30 cm$^{-1}$) -- as compared to water molecules in the bulk liquid. Mathies and Tauber [47b] speculate that these downshifts originate through the weakening of the H-O bonds due to the transfer of the excess electron density into the frontier orbitals of oxygen atoms. Similar downshifts, originating through this weakening, were observed in density functional theory and *ab initio* calculations of medium size water anion clusters. [49] By analogy, one might expect a similar pattern for $e_{am}^-$, where the solvation is by the N-H groups instead of the O-H groups. Despite an extensive search, no line shifts in the Raman spectra of liquid ammonia were found upon the addition of alkali metals. [50a] The only change observed is in the relative intensities of the combined $2\nu_4$ mode (asymmetric bend), the $\nu_1$ mode (symmetric stretch), and the $\nu_3$ mode (asymmetric stretch). In the IR spectra of dilute potassium solutions (< 5x10$^{-4}$ mol dm$^{-3}$), Rusch and Lugowski [50b] reported small downshifts of ca. 30 cm$^{-1}$ for all three of these modes. The difficulty of explaining these downshifts using the standard one-electron model has been recognized as early as in 1973, as seen from Jortner's remarks during the discussion at the end of refs. 50. On the other hand, it is even less clear whether Symon's model [3] of the solvent stabilized multimer anion of $e_{am}^-$ can account for the vibrational modes of this species.

As seen from this brief overview, further refinement of the cavity model is needed to account for the properties of the excess electron in aprotic liquids. It appears that ammonia, amines, and amides solvate the electron in a different fashion than water and alcohols. While we cannot presently provide a consistent many-electron model of electron solvation in *liquid* ammonia, a specific model of how this solvation *might* occur is given below. To this end, properties of small $am_n^-$ (n=2-8) cluster anions were examined theoretically. The ammonia molecules were arranged around the "cavity" in a fashion resembling the structure of $e_{am}^-$ obtained in one-electron models. [17-21] While such clusters are unrealistic models for gas phase multimer anions (that are unstable for $n < 30$), [22,23,25,26] we speculate that the resulting structures constitute the core of the excess electron in *liquid* ammonia: $e_{am}^-$ is indeed a *solvent stabilized radical multimer anion*. The cavity is formed due to the repulsion of negatively charged ammonia molecules sharing the excess electron density in the frontier orbitals of nitrogen atoms; only a fraction of the spin density resides inside the cavity. It is shown that the cluster anion model captures several observed features of $e^-_{am}$ that have not yet been accounted for theoretically, including the Knight shifts on $^{14}$N and $^{1}$H nuclei and the downshift of the stretch and bending modes. These calculations validate and elaborate the intuitive picture of electron solvation in ammonia suggested by the late Martyn Christian Raymond Symons, FRS (1925-2002), [3] to whose memory this paper is dedicated.

To reduce the length of the paper, figures with the designator "S" (e.g., Figure 1S) are placed in the Supplement.

**2. Computational Details.**

*2.1. Previous work.* The previously suggested *ab initio* models of the ammoniated electron [45,46] were based on self-consistent field (SCF) Hartree-Fock (HF) calculations for small (*n*=2-4), highly symmetrical cluster anions. The ammonia molecules where placed in such a way that either one of their N-H bonds looked straight towards the cavity center, which is designated "c" (B[bond]-orientation), or the lone pair of N looked away from this center (D[dipole]-orientation). Newton [45] examined the $D_{2d}$ symmetrical tetramer ammonia anion at the HF/4-31G level. All H-N-H angles were constrained to 113°, a ghost atom was added at the center, and dielectric continuum was placed beyond an arbitrary cutoff radius of 2.75 Å from this center. Isotropic hfcc for protons were estimated for the optimized structure; the *H 1s* spin density was ca -6.3x10$^{-4}$ $a_0^{-3}$ which corresponds to a hfcc of -1 G, i.e. $\Sigma_H A \approx$ -12 G, in a reasonable agreement with experiment. [28,29] Clark and Illing [46] used the HF method with 6-31+G$^*$ basis set for real atoms complemented by an extended set of *s*-functions at the cavity center in order to estimate the energetics of small cluster anions in the gas phase. Two dimer anions, a B-type one (with $C_{2h}$ symmetry) and a D-type one (with $D_{3d}$ symmetry) and a B-type trimer anion (with $C_{3h}$ symmetry) were examined. For optimized structures, the c-H distances were 2.5, 3.07, and 2.47 Å, respectively. The H-N-H angle (102-104°) was more acute than that for neutral ammonia molecules in the same HF model (ca. 107°). Clark and Illing, [46] like Newton [45] before them, were mainly interested in the energetics of these cluster anions. Still, their model suggested that the hfcc for the protons was small and negative. Clark and Illing, [46] however, observed that SCF calculations with split-valence basis sets typically result in unreliable estimates for these hfcc's.

*2.2. DFT models.* In this study, gas phase ammonia cluster anions were analyzed using density functional theory (DFT) models with BLYP functional (Becke's exchange functional [51] and the correlation functional of Lee, Yang, and Parr) [52] from Gaussian 98. [53] Several other functionals (e.g., the local spin density one) were also used, with fairly similar results. BLYP functional is most frequently used to estimate isotropic hfcc in radicals and radical ions, for which it typically yields accurate and reliable results. Unless specified otherwise, the basis set was a 6-31G split-valence double-$\zeta$ Gaussian basis set augmented with diffuse and polarized functions (6-31+G**). A ghost hydrogen or chlorine atom (i.e., floating-center basis functions) at the center or a cluster of such ghost atoms inside the cavity was added. It turned out that a single ghost atom was sufficient to provide the set of orbitals for filling the cavity. An increase in the number of ghost atoms did not significantly alter the results. The optimization of geometry was typically carried out using this basis set or a 6-31++G** basis set or an augmented Dunning's correlation consistent quadruple basis set (AUG-cc-pVQZ); [54] for the calculation of hfcc's and vibrational modes, the 6-31+G** basis set was used. We also carried out HF and second-order MØller-Plesset (MP2) perturbation theory [55] calculations using the same basis sets and obtained comparable results to those obtained using the DFT methods. This is reassuring as there are recognized pitfalls in using Becke's functionals for cluster anions (though these functionals are most frequently used to model such anions; [49] for a recent criticism of the DFT approach, see Herbert and Head-Gordon), [56] of which most important are overbinding for large basis sets and underbinding for small basis sets.

Some hfcc estimates in the DFT model were obtained using Barone's triple-$\zeta$ basis set with diffuse functions and an improved *s*-part (EPR-III), [57] but these estimates were reasonably close to those obtained using the 6-31+G** basis set and, therefore, are not reported except in section 3.5 and Table 2S. It appears that HF, MP, and DFT methods, regardless of the exact implementation and the choice of the basis set, yield the same basic electronic structure for ammonia anions. For this reason, only DFT models are considered henceforward.

All calculations discussed below were performed for gas phase cluster anions. We emphasize that these model species do not resemble at all the electron-trapping clusters observed in the gas phase (which, as suggested by the recent studies of small water cluster anions, [58,59] dipole-bind the electron at their surface).[58] The species of interest to us is the *core* of a much larger cluster anion that traps the electron in its interior or $e^-_{am}$ in the bulk solvent. Some calculations were carried out using the polarized continuum model of Tomasi et al.[60] in a similar fashion to Newton's model.[45] The main effect of introducing the continuum seems to be the tightening of the cluster. This contraction of the cavity leads, *inter alia*, to more negative hfcc's for the protons and greater hfcc's for the $^{14}N$ nuclei. The effect of the dielectric continuum on the partition of the highest occupied molecular orbital (HOMO) between the cavity and nitrogen atoms is small; most of the spin density remains in the *N* orbitals. Since such semicontinuum models depend on the arbitrary partitioning between the molecules and the "media" around them and do not capture the effect of hydrogen bonding between the solvent molecules in the first and the second solvation shells (that changes orientations of N-H bonds at the cavity wall), such calculations will not improve our knowledge of the structure. Hence we focus only on those aspects of the gas phase DFT models that are likely to relate to the observed properties of $e^-_{am}$ in liquid. For the same reason, we did not focus on the energetics of such gas phase clusters, as such energetics would bear little relation to that of the core anion in liquid ammonia. Our scope is limited only to the *structural* properties of this core anion; the energetics of solvation cannot be addressed using this crude approach.

Two basic geometries for ammonia clusters anions (*n*=2-8) were examined. In both of these geometries, the ammonia molecules were placed radially around the center (c). In the star-shaped B-type anions (e.g., anion **1** in Figure 1), one of the hydrogens ($H_a$) of each monomer looked towards the center (i.e., the $H_a$-c-N angle was constrained to 180°); the two other hydrogens ($H_b$) pointed away from this center, so that the c-$H_a$-N-H dihedral angles are ca. 124° (Table 1). In the D-type clusters (e.g., anion **2** in Figure 1), the $H_a$-c-N angles varied between 11° and 20° and the c-$H_a$-N-$H_b$ dihedral angles were ca.

54°: all three hydrogens pointed towards the center. In the gas phase, small D-type anions have lower energy than B-type anions, because this dipole orientation maximizes the attraction between H atoms and the electron in the cavity and minimizes the repulsion between the negatively charged nitrogens in the monomers. For $n$=2, 3, and 4 anions, a mirror plane symmetry was assumed, for n=4, 6 and 8 anions, and the nitrogens were arranged on the tetrahedral, octahedral, and cubic patterns, respectively. In these anions, all ammonia monomers had the same geometry and were placed at the same distance from the cavity center, so there were typically only three groups of magnetically equivalent nuclei (N, $H_a$, and $H_b$, see Figure 1).

Since the definition of what constitutes the cavity in a many-electron model of ammoniated electron is ambiguous, it is difficult to quantify the partition of the spin density between the cavity and the solvent molecules exactly. Examination of density maps for HOMO of the cluster anions suggests that the electron wavefunction inside the cavity and in the frontier orbitals of $N$ atoms have opposite signs, which makes it easy to distinguish these two contributions. Qualitatively, this partition can be assessed by examination of isodensity contour maps of spin-bearing HOMO of the anions (like those shown below in Figures 3 and 4). Typically, the diffuse, positive part of HOMO occupies 80-90% of the geometrical cavity at the density of +(0.01-0.03) e Å$^{-3}$ and less than 10% at the density of +(0.035-0.4) e Å$^{-3}$.

**3. Results.**

*3.1. The neutral monomer and some general trends for the ammonia anions.* In the BLYP/6-31+G** model of the $C_{3v}$ symmetrical ammonia molecule, the H-N bond is 1.025 Å and the H-N-H angle is 107.5°. These parameters may be compared with the crystallographic data for solid ammonia-I: 1.012 Å and 107.53°, respectively. [24] The calculated vibrational modes are in reasonable agreement with the experimental ones (Table 2). The symmetrical stretch ($v_1$) and asymmetrical bend ($v_4$) modes are least affected by the transfer from the gas to the liquid phase (in the liquid, the frequencies change significantly due to the hydrogen bonding, with a mean H-N…H distance of 2.357 Å); [24] these two modes are accurately estimated at the BLYP/6-31+G** level. In the anion clusters examined below, the excess electron density is partially transferred to

12.

nitrogen atoms, the N-H bonds are elongated by ca. 0.5% and the H-N-H angle is decreased from 107.5° to ca. 106° (in $am_8^-$) to ca. 102° (in $am_2^-$). The larger is the number of molecules sharing the negative charge, the smaller is the deviation from the neutral molecule geometry. For all cluster anions except for the cubic octamer, the D-type species have lower energy than the B-type ones. Such energetics are expected, as the preferred orientation of N-H bonds towards the cavity center observed in the PIMC models [17-21] of $e_{am}^-$ is due to the (i) electrostatic interaction of positively charged ammonia protons with the cavity electron and (ii) hydrogen bonding to ammonia molecules in the second solvation shell. Since in the gas-phase cluster anions (i) most of the excess electron density is on the nitrogen atoms and (ii) the second solvation shell is lacking, D-type orientation is favored energetically. We have examined the lowest-energy B-type anions nevertheless, as such anions may still be realized in *liquid* ammonia. The geometry of the clusters is summarized in Table 1 and the optimized structures are shown in Figures 1 through 6.

*3.2. Dimer.* The lowest energy B-type dimer anion **1** (Figure 1, *to the top*) has $C_{2h}$ symmetry; the D-type dimer anion **2** (Figure 1, *to the bottom*) has $D_{3d}$ symmetry and is ca. 150 meV more stable than anion **1**. Anion **1** is very tight (with $r(c - H_a) \approx 1.43$ Å vs. 2.9 Å in anion **2**) and, consequently, the spin density on N atoms is 50-70% greater than in medium size cluster anions. The N-H bond lengths, however, are fairly close to those in the neutral molecule: 1.026 and 1.030 Å in anion **1** and 1.026 Å in anion **2** (Table 1). The H-N-H angles are the smallest among all of the structures examined: 103-104° in anion **1** and 102.4° in anion **2**. The Mulliken charge and spin densities on nitrogen atoms are large (-1.52 and 0.62 for anion **1** and -1.68 and 0.77 for anion **2**, respectively (Table 1). Despite the large positive charge and spin density on $H_a$ protons in anion **1** (+0.44 and -0.088, respectively) vs. anion **2** (+0.37 and -0.064, respectively), the isotropic hfcc's on these protons are similar, -2.7 vs. -2.4 G, and isotropic hfcc's for $^{14}$N are also comparable, 27.7 vs. 22.9 G (Table 1). The anisotropic hyperfine coupling constants for the protons are ca. 2 times larger in anion **1** than in anion **2**, as may be expected from the small size of anion **1**. As the tensors of anisotropic hyperfine interaction are approximately axial (i.e., the principal values are $(+2T, -T, -T)$), these can be characterized by the largest

13.

principal value $2T$, which is ca. 6 G for $^1$H in anion **1** vs. 3.3 G in anion **2** (Table 1). The anisotropic hyperfine coupling for nucleus $X$ is given by $2T(X) \approx \gamma_e \gamma_X \langle (3\cos^2\alpha - 1)/r^3 \rangle$, where $\alpha$ is the angle between the $2p_z$ AO and the vector joining the nucleus and the unpaired electron and $r$ is the length of this vector; the angular bracket implies an average taken over the electronic wavefunction. The anisotropy of hyperfine interaction for $^{14}$N nuclei is small in both of these dimer anions ($2T_N < 1.2$ G), and it further decreases with size in larger anions (Table 1). The total isotropic hfcc on $^{14}$N and $^1$H nuclei are 55.4 G and -12 G in anion **1** and 45.7 G and -14.4 G in anion **2**, respectively. Notice that both the total atomic spin density and the isotropic hfcc's for the protons are negative. The charge density on the central ghost atom is fairly low (-0.1 for **1** and -0.15 for **2**), i.e. the charge and spin densities are mainly on the N atoms. Using more diffuse functions in the basis set does not change this partitioning appreciably. In the calculated Raman spectrum of anion **1**, the strongest bands are 3352 cm$^{-1}$ (B$_u$) and 3208 cm$^{-1}$ (A$_g$), there are also two weaker bands at 1635 and 3420 cm$^{-1}$, both of which are B$_g$ modes (Table 3). Thus, the asymmetric bend does not change and the stretch modes are strongly downshifted (Tables 2 and 3). While these dimer anions are poor reference models for $e^-_{am}$, many of the features observed for these dimer anions are also observed for larger cluster anions.

*3.3. Trimer.* The lowest energy B- and D- type trimer anions both have $C_{3h}$ symmetry (anions **3** and **4** in Figure 2, respectively); the former is ca. 160 meV more energetic than the latter. The H-N-H angles are still rather small (104° in anion **1** and 103° in anion **2**) and the N-H bonds are elongated (1.035 and 1.041 Å in anion **1** and 1.034 Å in anion **2**); see Table 1. The c-H$_a$ distance increases, as compared to the dimer anions, to 1.9 Å in anion **3** and 2.38 Å in anion **4**. Though the spin is equally divided between the three nitrogens (rather than two nitrogens in anions **1** and **2**), the total isotropic hfcc on $^{14}$N and $^1$H nuclei is not too different from that in the dimer anions (Table 1). The anisotropic hfcc for $^{14}$N are small ($2T_N < 1$ G) whereas those on the $^1$H$_a$ nuclei are still relatively large ($2T_H \approx 3.1$ G for anion **3** and 4.5 G for anion **4**). The isotropic hfcc's on the protons are small (Table 2) and the constants for $^{14}$N are large (+20.7 G for anion **3** and +17.7 G for anion **4**). While the geometry, spin and charge densities are sensitive to the choice of the basis set, especially when tight-binding sets (such as 4-31G and 6-31G) are used, the total hfcc's on $^{14}$N and $^{1}$H nuclei change only within 20%. The main effect of tight-binding is to make the anion smaller; the partitioning of the excess electron density between the three nitrogen atoms and the "cavity" changes relatively little. The same applies to other ammonia anions.

*3.4. Tetramer.* Both planar $C_{4h}$ symmetrical clusters and "tetrahedral" ($D_{2d}$) clusters (analogous to those studied by Newton) [45] were examined. The tightest of these clusters is a ring anion **5** shown in Figure 3 *(top left),* with extended hydrogen bonds (HN..H distance of 2.25 Å) between the monomers and c-$H_a$ separation of only 1.88 Å (Table 1). This cluster is so tight that the positive spin density from the cavity-filling part of the wavefunction compensates the negative contribution due to the bond polarization, and the isotropic hfcc on $H_a$ is slightly positive, +0.28 G. The resulting total isotropic hfcc on $^{14}$N and $^{1}$H nuclei are 80.7 G and -2.2 G, respectively. In larger anions, all hfcc on the protons are negative (Table 1). Anion **5** is more energetic by ca. 125 meV than star-shaped anion **6** (Figure 3, *top right*), which, in turn, is more energetic by ca. 150 meV than the $D_{2d}$ symmetric anion **7** (Figure 5, *top left*). Isodensity HOMO maps for $C_{4h}$ symmetric anions are shown in Figures 3 *(bottom right)* and 4 (both for anion **6**) and Figure 1S (for anion **8**). D-type anion **8** (with $C_{4h}$ symmetry; see Fig. 3, *bottom left*) and anion **9** (with $D_{2d}$ symmetry; see Figure 5, *top right*) are isoenergetic and by ca. 100 meV lower in energy than anion **7**. The total isotropic hfcc on $^{14}$N is +(50-65) G (i.e., the average hfcc for $^{14}$N is 13-16 G) and that for the protons is between -5.4 G (for anion **6**) and -12.3 G (for anion **9**); see Table 1. The latter parameter is systematically greater for D-type structures since there is no compensation of the negative hfcc constants (due to the interaction with the cavity electron) for inner $H_a$ protons in such anions. The anisotropic hyperfine interaction for $^{14}$N is weaker than in the dimer and trimer anions ($2T_N < 0.5$-$0.7$ G) and the protons are almost purely dipole coupled. Mulliken charge density on the N atoms is ca. -1.11 and the spin density is ca. 0.23-0.27. While the c-$H_a$ distances vary between 2.12 Å (in anion **6**) and 3.04 (in anion **9**), the mean isotropic hfcc's on $^{14}$N and $^{1}$H for these two tetramer anions are comparable (Table 1) reflecting the fact that the spin density is divided mainly among the ammonia molecules. The spin density *inside* the cavity is relatively small and there is a node at the center, as seen from the isodensity HOMO maps shown in Figures 4 and 5 (see also detailed maps in Figures 1S and 2S) For all of the anions examined, most of the spin density is in the frontier orbitals of N atoms.

For comparison to B-type dimer anion **1**, it is instructive to examine vibration modes in B-type anion **6**. The calculated Raman spectrum is dominated by the three $B_g$ bands at 1080, 1629, and 3201 cm$^{-1}$ that correspond to $\nu_2$, $\nu_4$, and $\nu_1$ modes of the neutral ammonia (Table 3). Once more, the asymmetric bend is relatively unchanged and the symmetric stretch is downshifted by ca. 170 cm$^{-1}$. For D-type anion **8**, the two most prominent Raman modes are at 1633 and 3248 cm$^{-1}$. The looser is the anion, the closer are the vibration frequencies to those of a neutral ammonia molecule.

*3.5. Hexamer and octamer.* Octahedral hexamer and cubic octamer anions are perhaps most instructive to examine because the coordination number of ammoniated electron, as suggested by PIMC models, [17] is six to nine.

In the $C_i$ symmetrical B-type anion **10** (Fig. 6, *top left*), the c-H$_a$ distances are 2.42 Å, and in D-type anion **11** (Fig. 6, *top right*), these distances are 3.3-3.4 Å (this anion has ca. 195 meV lower energy than anion **10**). The N-H bonds in the monomers are 1.03 Å (which is close to 1.025 Å in a neutral molecule) and the H-N-H angles are ca. 104.5° (vs. 107.5° in a neutral molecule). The total isotropic hfcc $\Sigma_N A$ on $^{14}$N is +50 G for anion **11** and +63.4 G for anion **10**, respectively; the total hfcc $\Sigma_H A$ on the protons is -6.8 G for anion **10** and -13.2 G for anion **11**, respectively (the tighter is the cluster anion, the more positive is $\Sigma_N A$). The smaller absolute values for $\Sigma_X A$ in D-type clusters vs. B-type clusters are also observed when larger triple-ζ sets (such as EPR-III) are used to calculate the hfcc constants. For anions 11 and 10 shown in Fig. 6, these calculations, for example, give +46.4 G and +56.5 G for $^{14}$N nuclei and -2.2 G and -6.3 G for the protons, respectively. Isodensity maps of singly occupied HOMO (SOMO) shown in Figure 3S indicate that the octahedral "cavity" is filled by the electron wavefunction (with a node at the center), but most of the spin density is divided between the frontier orbitals of N atoms (with Mulliken charge density of ca -1.0 and spin density of 0.15-0.17; see Table 1). To determine the effect of symmetry breaking on the hfcc's, the constraints were relaxed and several optimized structures were analyzed (see, e.g., Figure 4S). Despite the

16.

wide variation in the shape, bonding, and partitioning of the electron bonding between the nitrogens in the monomers, the total isotropic hfcc on $^{14}$N and $^{1}$H nuclei show surprisingly little variation. E.g., for the hexamer anion shown in Figure 4S, these constants are $\Sigma_N A \approx +54.2$ G and $\Sigma_H A \approx -12.1$ G, respectively. For D-type anions, the anisotropic hyperfine coupling constants are fairly close to those calculated in the point dipole approximation; e.g., for anion **11**, $2T$ for $^{14}$N and $^{1}$H nuclei are 0.64 and 1.4 G, respectively. The mean isotropic hfcc for $^{14}$N nuclei is typically around 10 G (vs. experimental 11±0.5 G for ammoniated F-center on MgO). [44]

Magnetic parameters for B-type anion **10** vs. the c-H$_a$ distance are given in Figure 8 and Table 1S (the structures were optimized for all other degrees of freedom). As this distance increases from 2 to 3 Å, $\Sigma_N A$ decreases from +78 to +52 G and $\Sigma_H A$ increases from -8.4 to -6.4 G. As seen from Figure 7, the hfcc's on outer protons change very slightly; the hfcc's on the inner protons change from -1 to -0.65 G. The total spin density on N atoms actually increases as the monomers move out, but the frontier N orbitals exhibit progressively more prominent *p*-character and the hfcc on $^{14}$N nuclei (that is sensitive only to the *s*-character) decreases. As seen from the comparison of Table 1 and Table 1S, the average isotropic and anisotropic hfcc's on $^{14}$N nuclei depend largely on the c-N distance rather than a specific arrangement of the monomers around the cavity or their number. This suggests that hfcc constants for $^{14}$N nuclei in the second solvation shell would not be negligible.

Table 2S demonstrates the effect of extending the basis set on the geometry and hyperfine coupling constants for B-type anion obtained using MP2 and BLYP methods. The corresponding SOMO maps for 6-31++G** basis set are shown in Figure 5S. These SOMO maps bear strong resemblance to the maps obtained using a tighter basis set, 6-31+G** (Fig. 3S). The MP2 method systematically yields smaller clusters as compared to the DFT methods, so the hfcc's (that were estimated using MP2 geometry and BLYP spin density) are accordingly larger, as can be surmised from Figure 7. BLYP calculations carried out using triple-ζ basis set EPR-III with extended *s*-type functions (Table 2S) indicate that parameters $\Sigma_N A$ obtained using double-ζ basis set 6-31+G* are 10-20% too high. Due to the smallness of hfcc's on the protons, this error is even larger for $\Sigma_H A$ (50-100%), for which there is comparable uncertainty in the experimental estimates for the (small) Knight shift on the ammonia protons. [28,29]

Cubic D- and B-type octamer anions have either $D_4$ or $C_{4h}$ symmetries. For undistorted cubic anions, the B-type anion has the lower energy; however, when elongation along the fourfold symmetry axis is allowed, the D-type anion has lower energy. In the $D_4$ symmetrical B-type anion **12** (Fig. 6, *bottom left*), the c-$H_a$ distance is 2.75 Å vs. 3.11 Å in the D-type anion **13** (Fig. 6, *bottom right*). For the latter, the parameters $\Sigma_N A$ and $\Sigma_H A$ are similar to those for D-type hexamer **11** (Table 1); for B-type anion **12**, these two parameters are +70.3 and -6.33 G, respectively. The anisotropy of hfcc on the nitrogens is very small ($2T_N < 0.5$ G) and for the protons the anisotropic coupling constants are comparable to those in the hexamer anions. Representative isodensity maps of SOMO for B-type anion **12** are shown in Figure 7. The spin density resides mainly on the N atoms but there is also a diffuse orbital filling the elongated cavity with a pronounced *p*-character. The calculated Raman spectrum of this anion (Table 3) is dominated by a single 3265-3267 cm$^{-1}$ band corresponding to the symmetric stretch (which is downshifted by ca. 100 cm$^{-1}$ from a free ammonia molecule). The asymmetrical bend modes are split into two branches; some of these are upshifted and some downshifted (Table 3). In B-type clusters, the downshifted modes are more prevalent, whereas in D-type clusters both of these two branches are present. It is difficult to predict from these data the overall effect of charge sharing on the combined $2v_4$ mode in the actual IR and Raman spectrum, but it is certain that this effect is relatively small and favors downshifts for N-H bonds pointing towards the cavity center.

Of all the multimer anions examined above only B-type octamer **12** seems to have a bound excited state, as suggested by time-dependent DFT calculations in the random phase approximation (for the BLYP/6-31+G** model). The transition is by 80% from the highest occupied to the lowest unoccupied MO; it is at 1.2 eV and it has an oscillator strength of 0.23. While it is encouraging that such a simple model predicts a bound-to-bound transition in the near IR, this lowest unoccupied MO consists of *N 2p* orbitals. It is unlikely that this transition has any bearing on the observed optical spectrum of $e_{am}^-$.

## 4. Discussion and Conclusions.

Our results suggest that the charge and spin density in small, highly symmetrical $am_n^-$ clusters preferentially resides on the frontier orbitals of N atoms of the cavity-forming ammonia molecules. Regardless of how many monomers are involved in the anion core, this partition results in the total isotropic hfcc of +(55-80) G on $^{14}$N and -(6-13) G on $^1$H. Although in these model calculations, almost no spin density resided in the "cavity", the total hfcc on $^{14}$N nuclei was still *lower* than the value of $\Sigma_N A \approx +110$ G obtained for $e_{am}^-$.[3,28] This apparent contradiction can be resolved by assuming, following Symons,[3] that in addition to the large (8-12 G) hfcc's for $^{14}$N nuclei in the 6-9 ammonia molecules forming the solvation cavity, there are small hfcc's for $^{14}$N nuclei in the 12-20 ammonia molecules that constitute the second solvation shell.

The most likely effect of this second solvation shell and the liquid behind it on the core anion would be the constriction of the solvation cavity. Such an effect can be obtained even in crude semicontinuum models.[45,60] In terms of the magnetic parameters, this constriction will increase $\Sigma_N A$ and make $\Sigma_H A$ more negative (see Table 1S and Figure 8). While it is presently impossible to assess quantitatively the effect of this constriction (see below), there is no obvious way in which the standard one-electron cavity model can explain the observed Knight shifts for $^{14}$N nuclei. Furthermore, only by transferring spin density into *N 2p* orbitals can one obtain negative hfcc's on the protons.

With the DFT calculations, there is always a possibility that unorthodox results are an artifact of tight-binding, and this problem is especially dire for the solvated electron as most of the density is outside the solvent molecules. Nothing in our results suggests that this is the case for ammonia anions. Introduction of additional sets of diffuse functions and ghost atoms does not have a large effect on how the spin density is divided between the cavity and ammonia molecules. Analogous DFT calculations for small water and methanol clusters do yield cavity electrons at the same level of modeling.[49, 57, 61] The fact that electron solvation in ammonia clusters is qualitatively different from that in water and methanol clusters has been suggested, albeit indirectly, by DFT calculations of sodium containing neutral clusters by Ferro and Allouche[62] and HF and MP calculations of Hashimoto and Morokuma. [63] In Na(H$_2$O)$_{7-10}$ and Na/methanol clusters, [62] the *Na 3s* electron is located far from the sodium nucleus; it can be regarded as a surface-trapped electron. By contrast, calculations of Na(NH$_3$)$_{6-11}$ clusters [62,63] indicated that the electron density was divided between the frontier orbitals of *N* atoms in ammonia molecules solvating Na, in a manner strikingly resembling the anions examined in section 3. This partitioning of the excess electron density among the ammonia molecules seems to be the natural result of DFT and *ab initio* modeling rather than an artifact of specific implementation of the model.

For hexamer and octamer anions (that most closely resemble the core of ammoniated electron), anisotropic hyperfine coupling constants are comparable with those estimated in the point-dipole approximation, suggesting inefficient nuclear relaxation due to such anisotropy, in agreement with experiment. [28,36,41,42] The calculations also suggest a small downshift for asymmetric bending modes (for B-type anions) and a relatively large downshift for symmetric stretching modes. Qualitatively, these results are in agreement with the experimental picture [47,48] for the *hydrated* electron (see the Introduction). The magnitude of the stretch mode downshift depends on the extent of delocalization of the electron between the ammonia molecules. Our analyses suggest that delocalization between 6-8 molecules already reduces this shift to 100-120 cm$^{-1}$ (Tables 2 and 3). As the NMR results indicate that even more electron delocalization should occur in $e^-_{am}$ (via the involvement of the second solvation shell, see above), the downshift will be reduced further. It seems entirely plausible that the resulting shift will be comparable to the experimental estimate of ca. 30 cm$^{-1}$. [50b] The DFT calculations specifically point to the population of frontier N orbitals as the cause of the observed downshift and account, within the limitations of the model, for the scale of this downshift.

The weak point of our model is its inability to address the optical properties and the energetics of $e^-_{am}$ that are precisely the properties that one-electron models tackle so well. This inability is not inherent to the DFT methodology: it is the consequence of limitations of our particular model, namely, the fact that we focused on small gas phase anions with fixed geometry. The real test would be a large-scale model in which DFT method is coupled to molecular dynamics, as in the recently published Car-Parrinello calculation of the hydrated electron.[64] In the absence of such a test, the model is incomplete, despite the suggestive indications that it captures the physics of the problem. We can, however, offer the following qualitative argument suggesting that the optical properties will be adequately accounted for in advanced DFT models. From the standpoint of one-electron cavity models,[5-11,17-23] $e^-_{am}$ is an electron inside a (nearly) spherical potential well. At the same level of idealization, the solvent-stabilized anion in ammonia can be viewed as an electron in a potential well that is shaped as a thin spherical layer (of nitrogen atoms in the first solvation shell), with some extension of the wavefunction towards the center of the cavity and towards the outside. Since both of these binding potentials are spherically symmetrical, the ground and excited states are $s$ and $p$ functions, in both of these two models. By suitable parameterization of these potentials and by allowing certain variation in these parameters, similar optical spectra can be obtained. Therefore, at the conceptual level it is very difficult to tell these two variants of the one-electron model apart.

We conclude that at some level of idealization, the many-electron and the one-electron cavity model of $e^-_{am}$ may look rather similar, provided that only a subset of the properties of ammoniated electron is taken into account. In one model, the cavity is formed due to the Coulomb repulsion between the solvent molecules sharing the negative charge; the excess electron resides on the frontier orbitals of $N$ atoms at the surface of the cavity. In another model, the electron fills up the interstitial void and forms its own cavity via Pauli repulsion of the solvent molecules. From the structural perspective, the end result (a cavity of a certain size) is the same. From the standpoint of the energetics, the end result (a spherically symmetrical potential well and the resulting absorption spectrum in the near IR) is also the same. Hence the success of one-electron models in explaining some properties of $e^-_{am}$.

In such a situation, the definitive test of the model is in its ability to reproduce the specific structural information, such as the spin density map given by the Knight shifts. For ammonia, this evidence points away from the cavity-filling one-electron model and towards, at the very least, the spherical-shell one-electron model, which has its natural explanation in the multimer radical anion model examined above.

21.

How general are these conclusions? Large Knight shifts for $^{14}$N and $^{13}$C nuclei for excess electron in dilute Na/methylamine solutions [31] suggest that electron solvation by the amino- and methyl- groups is qualitatively similar to that for the ammoniated electron. We have already suggested that electron solvation in alkanes [65] and acetonitrile [66] involves a solvent stabilized multimer anion with a fraction of the spin density transferred onto the frontier orbitals of *C* atoms in the methyl groups. It is very likely that a similar situation occurs in ethers, as such liquids also solvate the electron by their methyl and methylene groups. [68]

The mode of electron solvation in alkanes (that comprise the largest class of electron trapping liquids) can be addressed experimentally, in two different ways. First, it might be possible to determine Knight shifts for $^{13}$C nuclei in dilute Na/HMPA solutions. Catterall et al. [32] have already determined these shifts for $^{31}$P and $^{14}$N nuclei in HMPA; the small magnitude of these shifts and the fact that the absorption band of the electron is at 2.3 μm [33,34] suggest that the electron is solvated by methyl groups, with the polar P=O group looking away from the cavity (in a similar fashion to the solvated electron in acetonitrile). [66,67] Observation of the predicted large Knight shift on $^{13}$C nuclei in this liquid would provide direct evidence as to the occurrence of spin sharing by methyl groups. Alternatively, it might be possible to determine spin densities on $^{13}$C nuclei in $^{13}$CH$_3$ labeled glass-forming branched alkanes that are known to trap electrons below 77 K. [65,69] So far, the emphasis of the EPR and electron spin echo studies has been to determine *anisotropic* hfcc on the cavity protons. [69] Our models suggest that a measurement of isotropic hfcc on $^{13}$C nuclei would be a more direct probe of the mode of electron trapping. Equally important would be revisiting the EPR of hydrated electrons trapped in alkaline ices [70] since hyperfine couplings for $^{17}$O nuclei for the hydrated electron have never been determined and thus quantitative estimates as to the degree of penetration of the electron density on the frontier orbitals of oxygen atoms are lacking. Considerable downshifts for bending and stretching modes in the Raman spectra of the hydrated electron [47,48] and the prominent 180 nm absorption band in its optical spectrum [71] provide indirect evidence for partial occupation of these orbitals by the electron. The distinction between the cavity electron and the solvent stabilized multimer anion might be a matter of degree only. [72]

22.

**5. Acknowledgement.**

IAS thanks Professors F. T. Williams and B. J. Schwartz and Dr. J. F. Wishart for useful discussions and Drs. C. D. Jonah and M. C. Sauer, Jr. for careful reading of the manuscript and many suggestions. This work was supported by the Office of Science, Division of Chemical Sciences, US-DOE under contract number W-31-109-ENG-38.

*Supporting Information Available:* A single PDF file containing (1.) Figs. 1S to 5S with captions and (2.) Tables 1S and 2S. This material is available free of charge via the Internet at http://pubs.acs.org.

**References.**

**Figure captions.**

**Fig. 1**

B-type dimer anion **1** (the $C_{2h}$ symmetry; to the top) and D-type dimer anion **2** (the $D_{3d}$ symmetry; to the bottom); "c" is the inversion center. The two groups of magnetically equivalent protons are designated $H_a$ (inner protons) and $H_b$ (outer protons). Here and in other figures, the structures given have the lowest energy in the BLYP/6-31+G** density functional model. The structural and magnetic parameters for $^{14}N$ and $^{1}H$ nuclei for these anions are given in Table 1.

**Fig. 2**

Optimized geometry for the $C_{3h}$ symmetrical B-type trimer anion **3** (to the top) and D-type trimer anion **4** (to the bottom). The dashed lines point from the center of mass "c" towards $H_a$ protons (highlighted for anion **4**).

**Fig. 3**

Optimized geometry $C_{4h}$ symmetrical tetramer anions. To the top: Ring anion **5** (with hydrogen HN…H bonds between the monomers; *to the left*) and star-shaped B-type anion **6** (*to the right).* To the bottom: D-type anion **8** (to the left) and in-plane and out-of-plane surface density maps for singly occupied HOMO of anion **6** (*to the right*). The deep blue corresponds to -0.02 e Å$^{-3}$, the deep red corresponds to +0.02 e Å$^{-3}$.

**Fig. 4**

Isodensity surfaces for singly occupied HOMO of B-type tetramer anion **6** (the isodensity surfaces for anions **7** and **8** are given in Figures 2S and 1S, respectively). Scarlet is for positive density, violet is for negative density. The surfaces correspond to (a) ±0.01, (b) ±0.02, (c) ±0.03, and (d) ±0.04 e Å$^{-3}$. The cross marks the center of mass. Most of the spin density is in the frontier orbitals of *N* atoms; the diffuse positive wavefunction that envelops the anion has a node at the center.

**Fig. 5**

Optimized geometry $D_{2d}$ symmetrical tetramer anions. To the top: B-type anion **7** (to the left) and D-type anion **9** (to the right). To the bottom: Isodensity surfaces for singly occupied HOMO of anion **7** (two more such surfaces are shown in Figures 2S). The same convention as in Fig. 4. The surfaces correspond to (a) ±0.01 and (b) ±0.03 e Å$^{-3}$.

**Fig. 6**

Optimized geometry "octahedral" hexamer and "cubic" octamer anions. To the top: The $C_i$ symmetrical B-type hexamer anion **10** (to the left) and D-type anion **11** (to the right). These structures may be obtained from the $C_{4h}$ symmetrical tetramer anions by placing two more ammonia monomers along the fourfold rotation axis. To the top: The $D_4$ symmetrical B-type octamer anion **12** (*to the left*) and D-type anion **13** (*to the right*). For both of these octamer anions, $H_a$ protons are highlighted.

**Fig. 7**

Isodensity surfaces for singly occupied HOMO of B-type cubic octamer anion **12** (the same conventions as in Fig. 4). The surfaces correspond to (a) ±0.01, (b) ±0.015, (c) ±0.02, and (d) ±0.03 e Å$^{-3}$.

**Fig. 8**

(a) Relative binding energy, (b) mean isotropic hfcc's on $^{14}$N *(filled squares, to the left)*, $^1H_a$ (empty circles, to the right), and $^1H_b$ *(empty triangles, to the right)* nuclei, and (c) total hfcc on $^{14}$N *(filled squares, to the left)* and $^1$H *(empty squares, to the rght)* nuclei for B-type $C_i$ symmetrical hexamer anion **10** (Fig. 6) as a function of c-$H_a$ distance. See Table 1S for structural and magnetic parameters.

**Table 1.**

Geometry, atomic spin and charge densities, and magnetic parameters for model $am_n^-$ cluster anions (BLYP/6-31G+**).

| anion, $am_n^-$ | 1 | 2 | 3 | 4 | 5 | 6 | 7 |
|---|---|---|---|---|---|---|---|
| $n$; type | 2, B | 2, D | 3, B | 3, D | 4, ring | 4, B | 4, B |
| symmetry | $C_{2h}$ | $D_{3d}$ | $C_{3h}$ | $C_{3h}$ | $C_{4h}$ | $C_{4h}$ | $D_{2d}$ |
| r(c-$H_a$) | 1.436 | 2.189 | 1.903 | 2.378 | 1.88 | 2.358 | 2.122 |
| r(c-N) | 2.471 | 2.432 | 2.944 | 2.876 | 2.317 | 3.396 | 3.16 |
| r(N-$H_a$) | 1.034 | 1.026 | 1.041 | 1.037 | 1.043 | 1.038 | 1.038 |
| r(N-$H_b$) | 1.026 | 1.026 | 1.035 | 1.034 | 1.032 | 1.032 | 1.031 |
| a($H_a$-c-N) | 0 | 24.9 | 0 | 20 | 26.2 | 0 | 0 |
| a($H_a$-N-$H_b$) | 103 | 102.4 | 103.7 | 102.7 | 106.3 | 104.1 | 104.4 |
| a($H_b$-N-$H_b$) | 103.8 | 102.4 | 104.2 | 102.8 | 104.8 | 104.5 | 105.3 |
| d(c-$H_a$-N-$H_b$) | 53.9 | 53 | 125.6 | 53.2 | 124.4 | 125.4 | 124.9 |
| A($^{14}$N) | 27.7 | 22.9 | 20.7 | 17.7 | 20.2 | 16 | 14.5 |
| -A($^1H_a$) | 2.7 | 2.4 | 0.85 | 1.3 | -0.28 | 0.77 | 0.91 |
| -A($^1H_b$) | 1.65 | 2.4 | 0.55 | 1.35 | 0.41 | 0.29 | 0.52[a] |
| $\Sigma_N A$ | 55.4 | 45.7 | 62.1 | 53.1 | 80.7 | 64 | 58.2 |
| $-\Sigma_H A$ | 12 | 14.4 | 11.7 | 12 | 2.17 | 5.4 | 7.8 |
| 2T($^{14}$N) | 1.05 | 1.24 | 0.75 | 1.0 | 0.1 | 0.5 | 0.7 |
| 2T($^1H_a$) | 6.1 | 3.3 | 4.5 | 3.1 | 1.03 | 3.4 | 3.75 |
| 2T($^1H_b$) | 8.6 | 3.3 | 2.1 | 2.3 | 1.42 | 1.5 | 1.7 |
| $-\rho_c$(N) | 1.52 | 1.68 | 1.23 | 1.32 | 1.23 | 1.1 | 1.12 |
| $\rho_c(H_a)$ | 0.44 | 0.37 | 0.35 | 0.33 | 0.42 | 0.31 | 0.32 |
| $\rho_c(H_b)$ | 0.31 | 0.37 | 0.29 | 0.32 | 0.28 | 0.285 | 0.28 |
| $\rho_s$(N) | 0.62 | 0.77 | 0.35 | 0.43 | 0.30 | 0.23 | 0.24 |
| $\rho_s(H_a)$, x100 | -8.7 | -6.4 | -1.2 | -2 | -3 | 1 | 0.7 |
| $\rho_s(H_b)$, x100 | -4 | -6.4 | -1.5 | -2.5 | -1.2 | -0.8 | 0.75 |

*continued*

| anion, $am_n^-$ | 8 | 9 | 10 | 11 | 12 | 13 |
|---|---|---|---|---|---|---|
| $n$; type | 4, D | 4, D | 6, B | 6, D | 8, B | 8, D |
| symmetry | $C_{4h}$ | $D_{2d}$ | $C_i$ | $C_i$ | $D_4$ | $D_4$ |
| r(c-$H_a$) | 2.774 | 3.136 | 2.418 | 3.287 | 2.748 | 3.113 |
| r(c-N) | 3.417 | 3.368 | 3.453 | 3.667 | 3.78 | 3.866 |
| r(N-$H_a$) | 1.035 | 1.031 | 1.035 | 1.031 | 1.030 | 1.03 |
| r(N-$H_b$) | 1.032 | 1.033 | 1.030 | 1.031 | 1.030 | 1.029 |
| a($H_a$-c-N) | 15.2 | 17.8 | 0 | 15.9 | 0 | 11.6 |
| a($H_a$-N-$H_b$) | 103.9 | 103.6 | 104.9 | 104.5 | 104.3 | 105.7 |
| a($H_b$-N-$H_b$) | 103.9 | 103.1 | 105.7 | 104.6 | 105.9 | 105.9 |
| d(c-$H_a$-N-$H_b$) | 54.2 | 53.7 | 124.4 | 54.8 | 125.3 | 55 |
| A($^{14}$N) | 13.5 | 12.8 | 10.6[a] | 8.3[a] | 8.8 | 7.4 |
| -A($^1H_a$) | 0.99 | 0.91[a] | 0.74[a] | 0.75[a] | 0.66 | 0.78 |
| -A($^1H_b$) | 0.95 | 1.09[a] | 0.2[a] | 0.73[a] | 0.13[a] | 0.44[a] |
| $\Sigma_N A$ | 11.6 | 51.3 | 63.4 | 49.7 | 70.3 | 57 |
| $-\Sigma_H A$ | 11.6 | 12.3 | 6.8 | 13.2 | 6.3 | 13.1 |
| 2T($^{14}$N) | 0.8 | 0.78 | 0.5 | 0.64 | 0.38 | 0.52 |
| 2T($^1H_a$) | 2.7 | 1.64 | 3.0 | 1.4 | 2.6 | 1.6 |
| 2T($^1H_b$) | 1.7 | 2.2 | 1.3 | 1.3 | 0.9 | 1.0 |
| $-\rho_c$(N) | 1.17 | 1.17 | 0.97 | 1.05 | 0.97 | 1.0 |
| $\rho_c$($H_a$) | 0.32 | 0.3 | 0.29 | 0.30 | 0.28 | 0.3 |
| $\rho_c$($H_b$) | 0.30 | 0.31 | 0.28 | 0.29 | 0.29 | 0.29 |
| $\rho_s$(N) | 0.28 | 0.29 | 0.12[a] | 0.17 | 0.09 | 0.13 |
| $\rho_s$($H_a$), x100 | -0.95 | -1.3 | 2.6 | -0.75[a] | 2.4 | -0.6 |
| $\rho_s$($H_b$), x100 | -1.2 | -1 | ≈0 | -0.2[a] | 0.1 | -0.2 |

Bond distances (r) are in Å, bond (a) and dihedral (d) angles in °, isotropic hfcc's (A) for the given nuclei, sum total isotropic hfcc ($\Sigma A$) for $^{14}$N and $^1$H and maximum principal values of the tensor for anisotropic hyperfine interaction (*2T*) in Gauss, Mulliken atomic spin ($\rho_s$) and charge ($\rho_c$) densities in e Å$^{-3}$. Symbol "c" stands for the cavity center; *n* is the number of ammonia molecules; B is for bond- and D- is for dipole oriented cluster anions. (a) average value.

**Table 2.**

Normal vibrational modes of ammonia molecules.

| mode | Molecule (gas phase) | Liquid Ammonia [b] | Molecule Calc.[c] |
|---|---|---|---|
| $\nu_1$ (*symm. stretch*, $A_1$) [a] | *3334.2* | 3285 | 3370 |
| $\nu_2$ (*symm. bend*, $A_1$) | 932 & 968 *934 & 964.3* | 1035-1066 | 993 |
| $\nu_3$ (*asymm. stretch*, E) | 3414 | 3375 | 3507.6 |
| $\nu_4$ (*asymm. bend*, E) | 1627.5 | 1632 | 1636 |

The frequencies are given in cm$^{-1}$. Raman shifts are given in italics.

a) double (inversion) bands for $C_{3v}$ symmetrical molecule; the representations are given in parentheses.

b) center band positions from ref. 50; only $\nu_1$, $\nu_3$, and $2\nu_4$ bands are observed in IR and Raman spectra of liquid ammonia.

c) BLYP/6-31+G** calculation

**Table 3.**

Most prominent Raman-active normal modes for selected ammonia cluster anions (BLYP/6-31+G**, optimized geometry).

| mode | 1 | 6 | 9 | 10 | 11 | 12 | 13 |
|---|---|---|---|---|---|---|---|
| $n$, type | 2, B | 4, B ($C_{4h}$) | 4, D ($D_{2d}$) | 6, B | 6, D | 8, B | 8, D |
| $\nu_2$ [a] symm. bend | 1046 1109 | 1081-3 | 1171 | 1074 *(43)* 1087 *(14)* 1090 *(42)* | 1142 | 1110 *(49)* 1120 *(42)* | 1057 *(10)* 1069 *(24)* 1079 *(66)* |
| $\nu_4$ asymm. bend | 1632-5 | 1629 | 1631-7 *(41)* 1644-8 *(50)* 1651 *(27)* | 1618 *(26)* 1621 *(10)* 1629 *(45)* | 1628 *(18)* 1637 *(10)* 1639 *(56)* | 1624 *(20)* 1627 *(8)* 1638 *(65)* | 1624-7 *(16)* 1634 *(21)* 1639 *(25)* 1647 *(38)* |
| $\nu_1$ symm. stretch | 3208 | 3201 | 3251-3263 | 3245 *(41)* 3250 *(46)* | 3286-7 | 3262-5 | 3294-7 |

The frequencies are given in cm$^{-1}$. Raman shifts are given in italics. The same notations as as in Table 1 for the anions. The fraction of the total intensity in a given band (in %) is given in parentheses (in italics). (a) For neutral ammonia clusters, this frequency is blue shifted to 1000-1020 cm$^{-1}$.

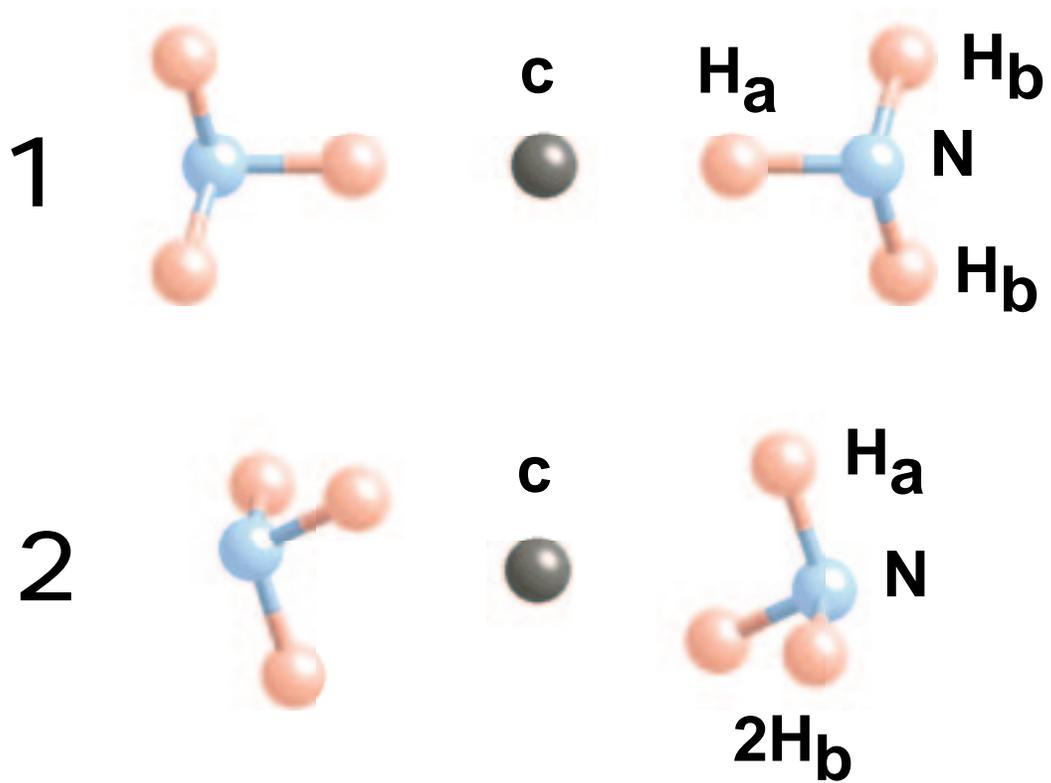

**Figure 1; Shkrob**

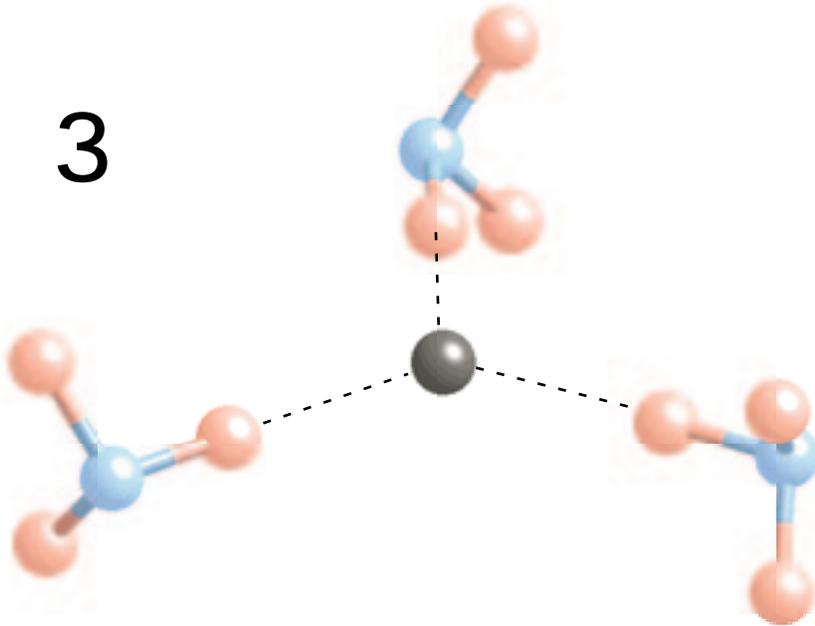

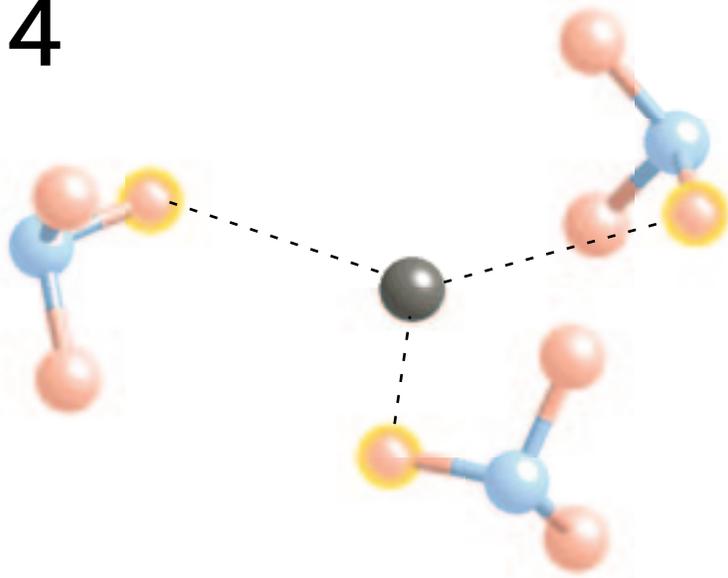

**Figure 2; Shkrob**

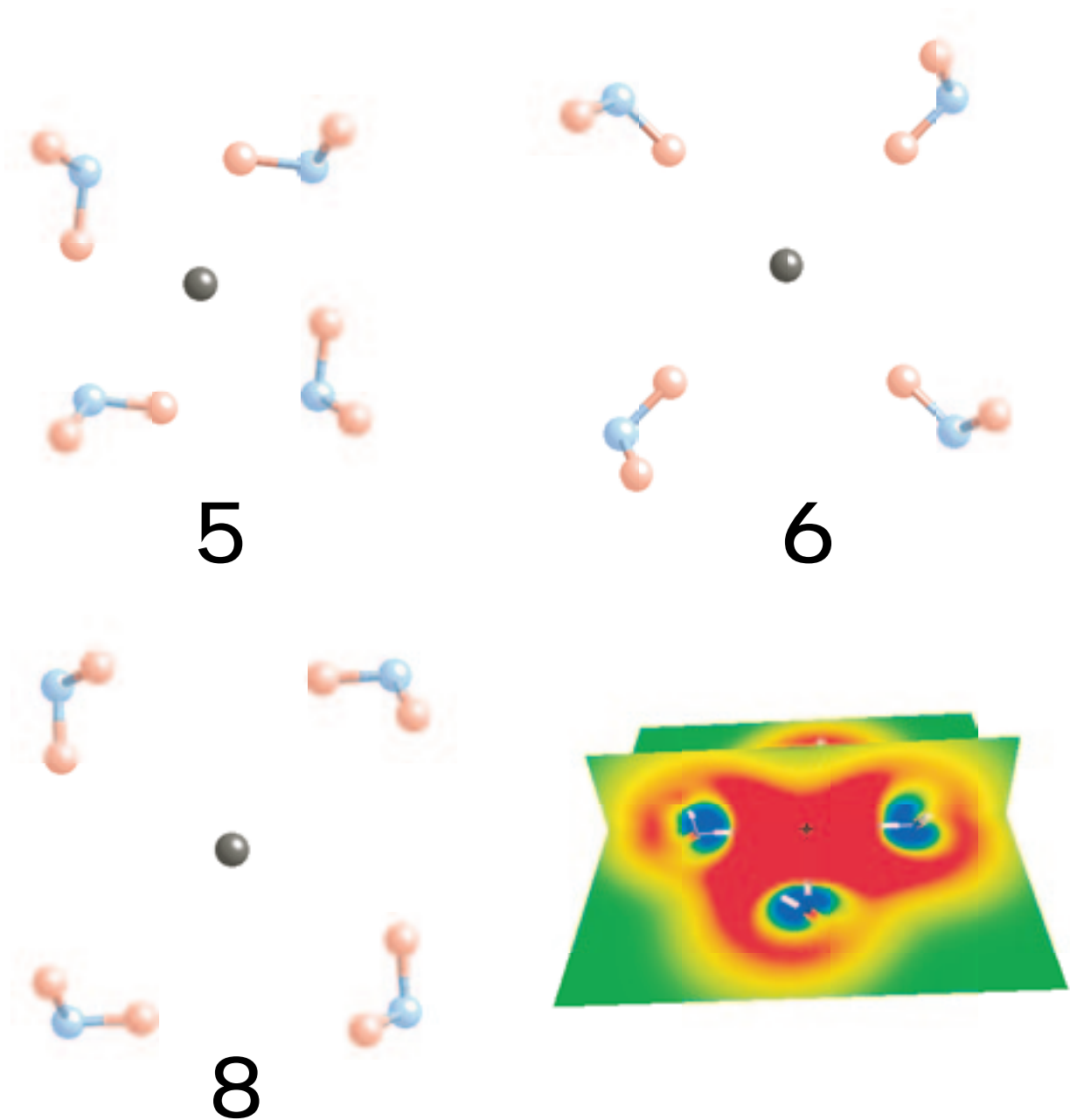

Figure 3; Shkrob

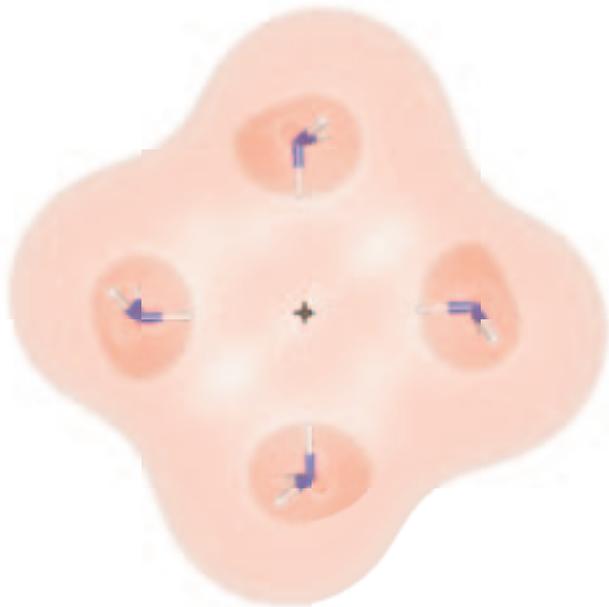
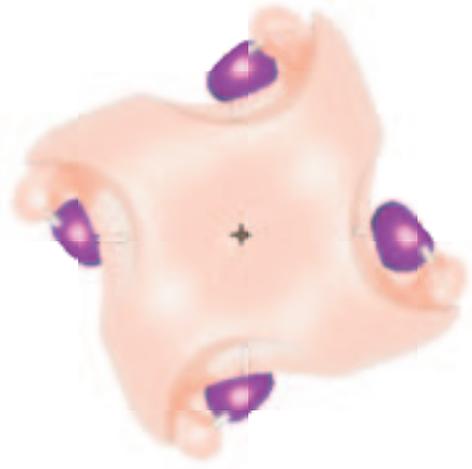

**(a)**        **(b)**

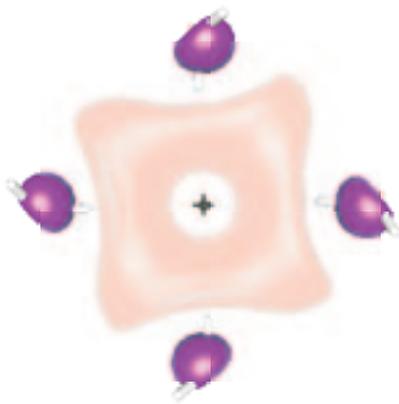
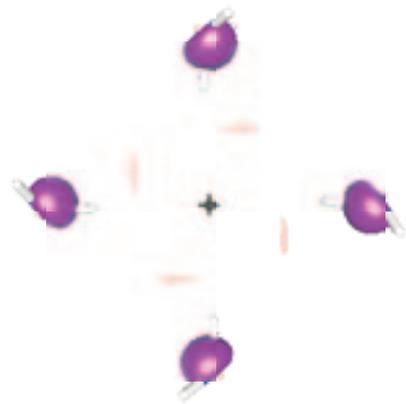

**(c)**        **(d)**

**Figure 4; Shkrob**

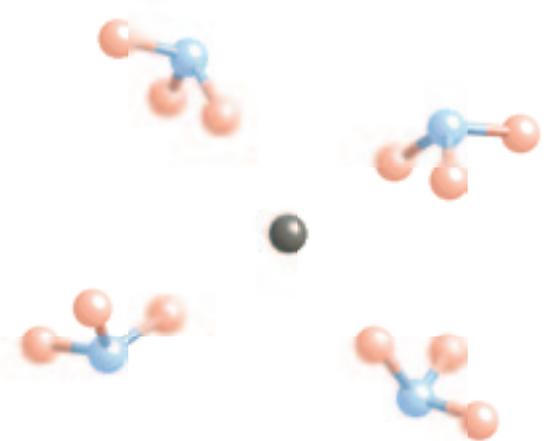
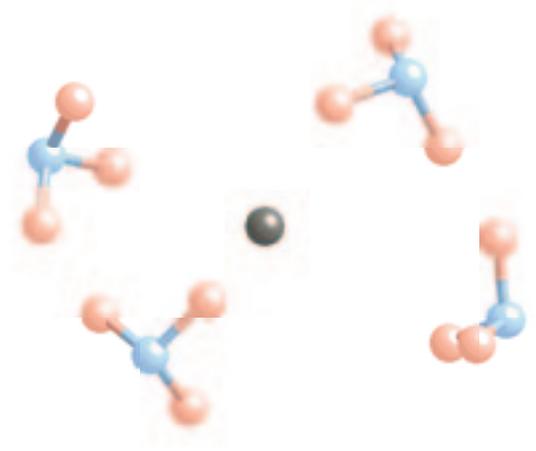

**7**          **9**

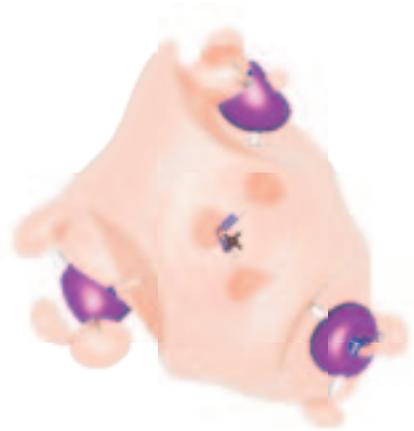
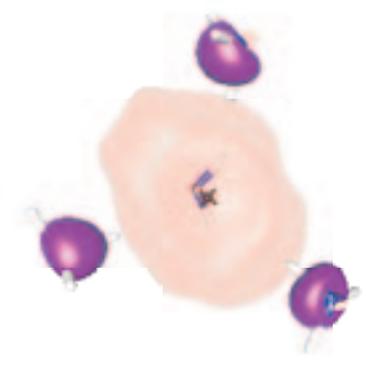

(a)          (b)

**Figure 5; Shkrob**

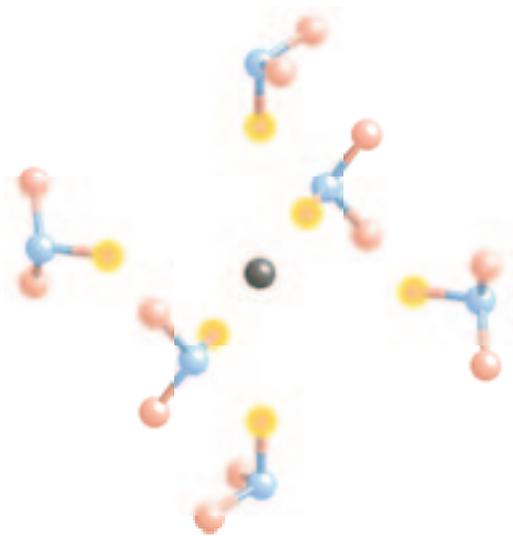
**10**

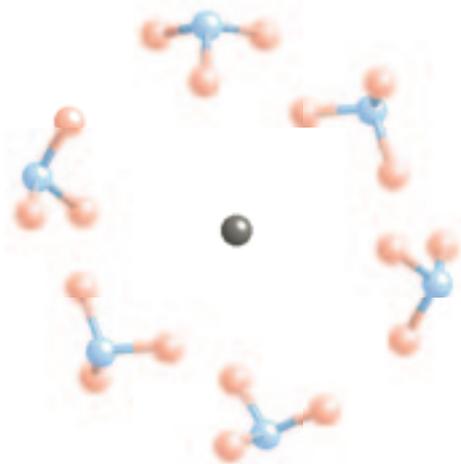
**11**

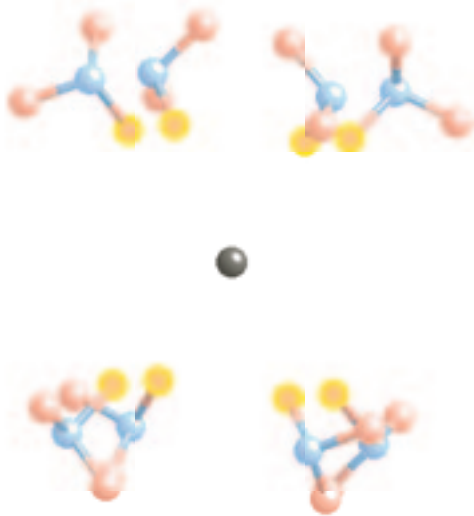
**12**

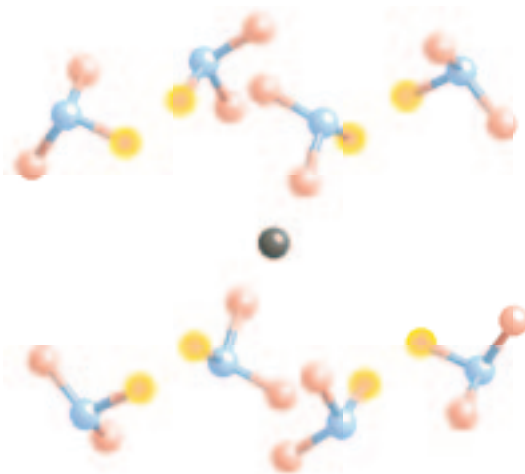
**13**

**Figure 6; Shkrob**

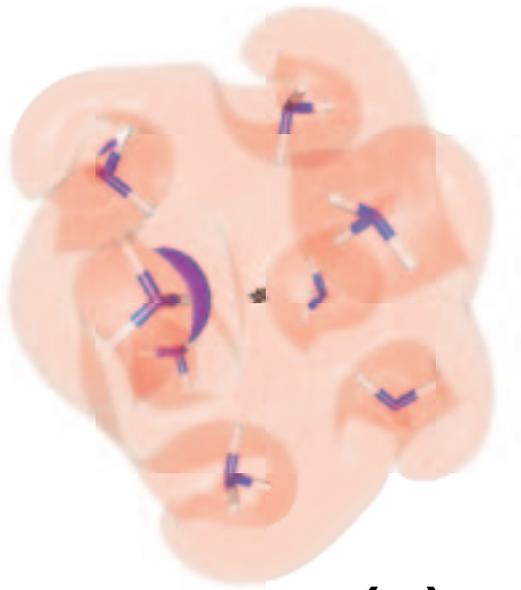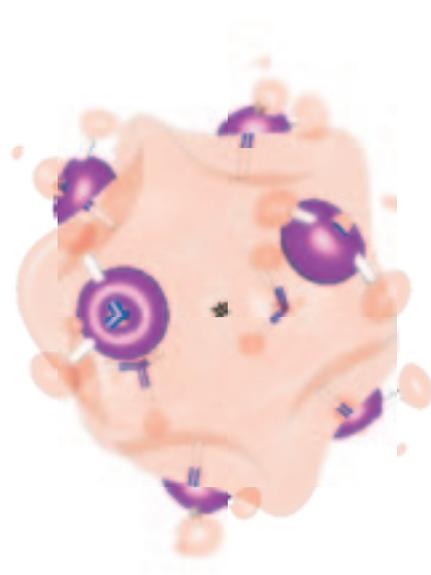

(a) (b)

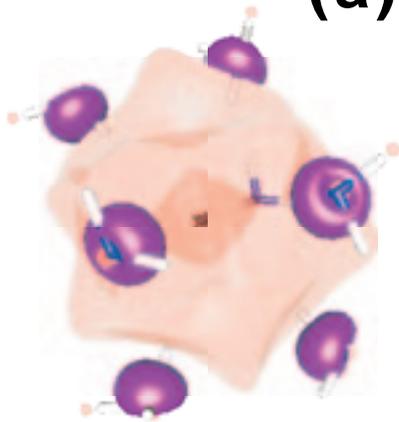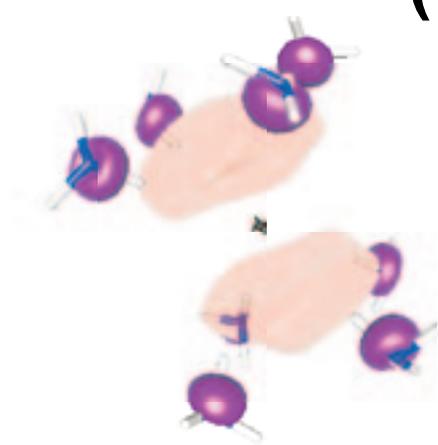

(c) (d)

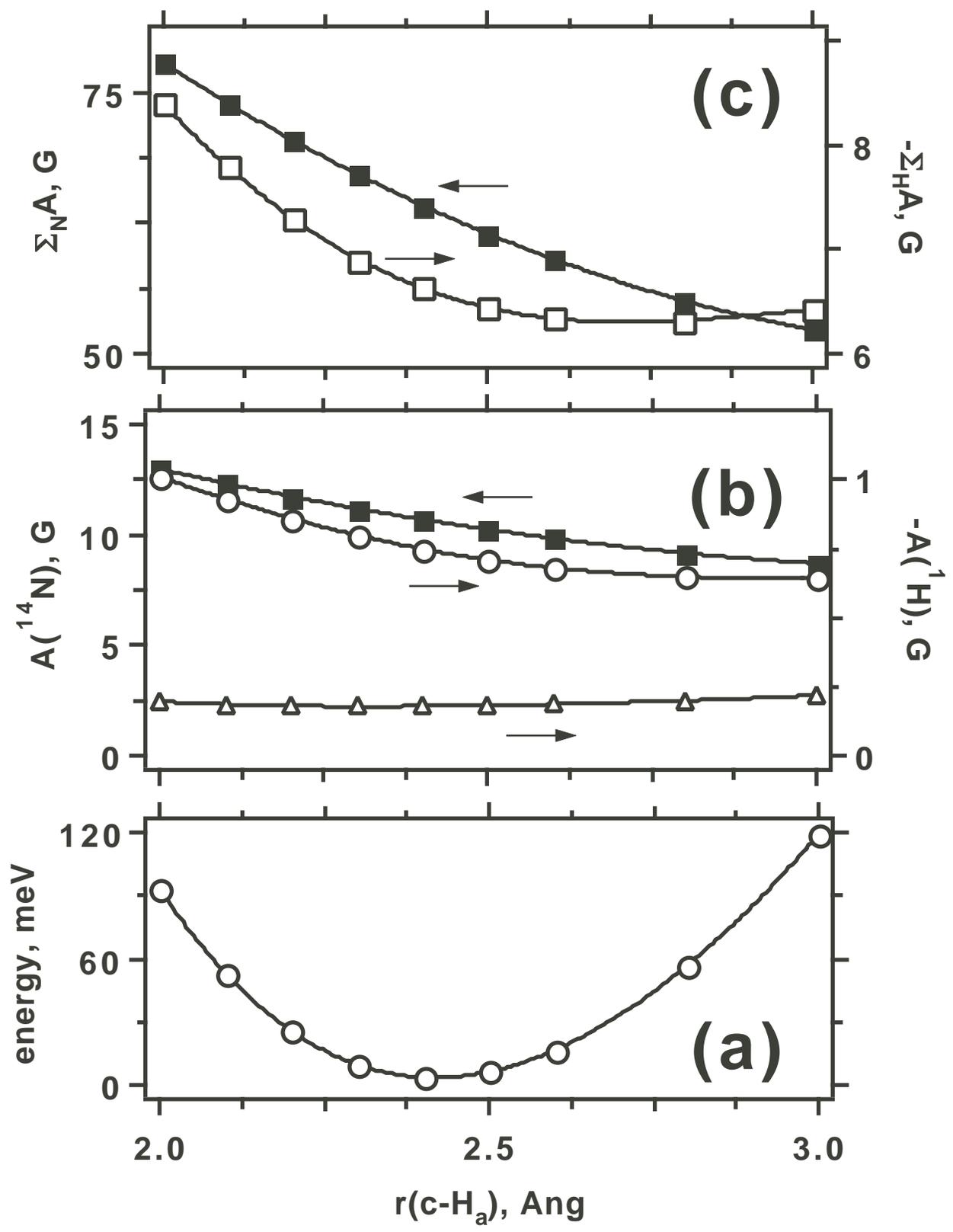

Figure 8; Shkrob

**Ammoniated electron as a solvent stabilized multimer radical anion.**

*Ilya A. Shkrob*

*Radiation and Photochemistry Group, Chemistry Division, Argonne National Laboratory, 9700 South Cass Avenue, Argonne, Illinois 60439*

*Tel* 630-2529516, *FAX* 630-2524993, *e-mail:* shkrob@anl.gov.

# Supporting Information.

**(1). Figure captions (1S to 5S).**

**Fig. 1S.**

The same as Fig. 4, for $C_{4h}$ symmetrical D-type tetramer anion **8**.

**Fig. 2S.**

The same as Fig. 4, for $D_{2d}$ symmetrical B-type tetramer anion **7**. The surfaces are densities corresponding to (a) ±0.01, (b) ±0.025, (c) ±0.03, and (d) ±0.035 e Å$^{-3}$.

**Fig. 3S.**

The same as Fig. 2S, for $C_i$ symmetrical B-type hexamer anion **10**.

**Fig. 4S.**

One of the hexamer anions with a loose octahedral pattern obtained with relaxed constraints. Some ammonia molecules are B-oriented, some are D-oriented. Despite the absence of order, the total hfcc's on $^{14}$N and $^{1}$H nuclei show relatively little variation in such loose clusters.

**Fig. 5S.**

The same as Fig. 3S for "octahedral" hexamer anions modeled using (a) MP2 and (b) BLYP methods with 6-31++G** basis set (a chlorine ghost atom was placed at the center). Isodensity contour maps of SOMO for optimized geometry hexamer anions are shown. The densities are, from top to bottom: (a) ±0.017, ±0.02, ±0.03, and ±0.0335, and

(b) ±0.016, +0.02, ±0.03, and ±0.035 e Å$^{-3}$. The geometry and magnetic parameters are given in Table 2S.

**(2). Table 1S.**

Geometry, atomic spin and charge densities, and magnetic parameters for B-type $am_6^-$ cluster anions as a function of c-$H_a$ distance (BLYP/6-31G+**).

| r(c-$H_a$) | 2.0 | 2.1 | 2.2 | 2.3 | 2.4 | 2.5 | 2.6 | 2.8 | 3.0 |
|---|---|---|---|---|---|---|---|---|---|
| r(N-$H_a$) | 1.039 | 1.038 | 1.037 | 1.036 | 1.035 | 1.034 | 1.033 | 1.032 | 1.031 |
| d(c-$H_a$-N-$H_b$) | 124.6 | 124.5 | 124.5 | 124.4 | 124.4 | 124.4 | 124.4 | 124.4 | 124.4 |
| A($^{14}$N) | 13.0 | 12.3 | 11.8 | 11.2 | 10.7 | 10.3 | 9.85 | 9.2 | 8.7 |
| -A($^1H_a$) | 1.01 | 0.93 | 0.86 | 0.8 | 0.75 | 0.71 | 0.68 | 0.65 | 0.65 |
| -A($^1H_b$) | 0.197 | 0.2 | 0.19 | 0.18 | 0.18 | 0.18 | 0.19 | 0.2 | 0.21 |
| $\Sigma_N A$ | 77.9 | 74 | 70.5 | 67.2 | 64.2 | 61.5 | 59.1 | 55.1 | 52.3 |
| $-\Sigma_H A$ | 8.4 | 7.8 | 7.3 | 6.9 | 6.6 | 6.4 | 6.3 | 6.3 | 6.4 |
| 2T($^{14}$N) | 0.65 | 0.58 | 0.53 | 0.49 | 0.45 | 0.44 | 0.37 | 0.33 | 0.31 |
| 2T($^1H_a$) | 4.3 | 3.9 | 3.6 | 3.3 | 2.8 | 2.6 | 2.4 | 2.0 | 1.74 |
| 2T($^1H_b$) | 1.5 | 1.4 | 1.3 | 1.2 | 1.1 | 1.0 | 1.0 | 0.85 | 0.73 |
| $-\rho_c(N)$ | 0.98 | 0.98 | 0.99 | 0.99 | 1.0 | 1.0 | 1.01 | 1.02 | 1.03 |
| $\rho_c(H_a)$ | 0.287 | 0.283 | 0.286 | 0.288 | 0.291 | 0.294 | 0.297 | 0.303 | 0.308 |
| $\rho_s(N)$, x10 | 1.04 | 1.07 | 1.1 | 1.17 | 1.23 | 1.29 | 1.35 | 1.47 | 1.6 |
| $\rho_s(H_a)$, x$10^2$ | 3.6 | 3.5 | 3.3 | 3.07 | 2.75 | 2.4 | 2.0 | 1.3 | 0.53 |
| $\rho_s(H_b)$, x$10^3$ | 2.8 | 2.0 | 1.3 | 0.4 | -0.5 | -1.3 | -2.2 | -3.9 | -5.5 |
| $\Delta E$, meV | 93 | 53 | 26 | 10 | 4 | 6 | 17 | 57 | 119 |

See the legend for Table 1 for units and notations used. $\Delta E$ is the energy relative to the optimized anion. a($H_a$-N-$H_b$) is ca. $104.8^0$, r(N-$H_b$) is ca. 1.03 Å, and $\rho_c(H_b)$ is ca. 0.28 e Å$^{-3}$ for all of these structures. Mean isotropic and anisotropic hfcc's for $^{14}$N and $^1$H nuclei are in Gauss (1 G = $10^{-4}$ T).

**Table 2S.**

Geometry, atomic spin and charge densities, and magnetic parameters for $C_i$ symmetrical B-type ("octahedral") ammonia anions.

| method | BLYP | MP2 | BLYP | MP2 |
|---|---|---|---|---|
| basis set for optimization | 6-31+G* | 6-31+G* | 6-31++G** | 6-31++G** |
| r(c-H$_a$) | 2.418 | 2.370 | 2.501 | 2.333 |
| r(c-N) | 3.452 | 3.390 | 3.533 | 3.351 |
| r(N-H$_b$) | 1.030 | 1.015 | 1.030 | 1.016 |
| a(H$_a$-N-H$_b$) | 104.9 | 105.5 | 105.5 | 105.9 |
| d(c-H$_a$-N-H$_b$) | 124.4 | 123.8 | 123.9 | 123.4 |
| A($^{14}$N) | 10.6 (8.0[b], 9.4[c]) | 7.2 (11.9[a], 9.6[c]) | 7.8 (9.2[c]) | 5.1 (8.4[b], 8.55[c]) |
| -A($^1$H$_a$) | 0.74 (0.31[b], 0.42[c]) | 4.8 (0.92[a], 0.4[c]) | 0.39 (0.38[c]) | 2.2 (0.4[b], 0.47[c]) |
| -A($^1$H$_c$) | 0.18 (0.05[b], 0.03[c]) | 0.92 (0.21[a], 0.02[c]) | 0.08 (0.05[c]) | 0.66 (0.04[b], ≈0[c]) |
| $\Sigma_N A$ | 63.8 (48.2[b], 56.5[c]) | 43 (71.3[a], 57.6[c]) | 46.7 (55.1[c]) | 30.8 (50.5[b], 59.1[c]) |
| $-\Sigma_H A$ | 6.6 (2.5[b], 2.2[c]) | 39.6 (8[a], 3[c]) | 3.3 (1.7[c]) | 21.3 (2.9[b], 2.8[c]) |
| 2T($^{14}$N) | 0.51 (0.32[b], 0.36[c]) | 0.5 (0.51[a], 0.4[c]) | 0.27 (0.3[c]) | 0.17 (0.37[b], 0.4[c]) |
| 2T($^1$H$_a$) | 3.2 (2.0[b], 2.1[c]) | 4.0 (2.75[a], 2.0[c]) | 1.9 (2.0[c]) | 1.8 (1.9[b], 2.2[c]) |
| 2T($^1$H$_c$) | 1.3 (0.9[b], 1.2[c]) | 1.1 (0.98[a], 1.2[c]) | 0.8 (1.1[c]) | 0.6 (0.9[b], 1.0[c]) |

Bond distances (r) are in Å, bond (a) and dihedral (d) angles are in ° (optimized geometry); average isotropic hfcc's (A) for the given nuclei, sum total isotropic hfcc ($\Sigma A$) for $^{14}$N and $^1$H and maximum principal values of the tensor for anisotropic hyperfine interaction (2T) are in Gauss. Symbol "c" stands for the cavity center (at which a ghost chlorine atom was placed). Calculated using BLYP method with (a) 6-31+G** double-ζ basis set; (b) 6-31++G** double-ζ basis set; (c) EPR-III triple-ζ basis set with improved s-part. For neutral $C_{3v}$ symmetrical ammonia molecule, r(N-H)=1.025 and 1.012 Å and a(H-N-H)=107.5° and 108°, in the BLYP and MP2 models with 6-31++G** basis set, respectively.

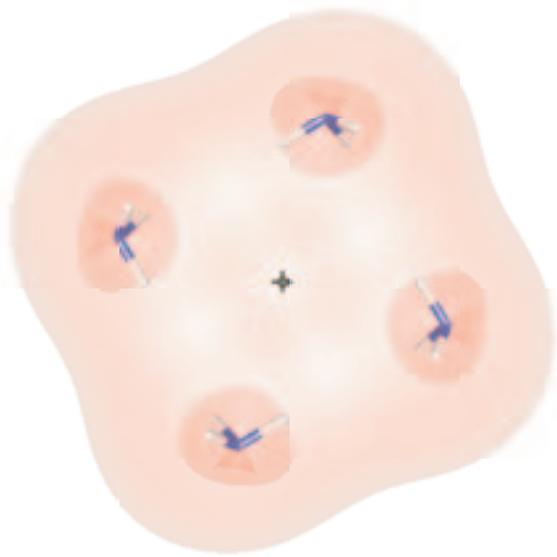
**(a)**

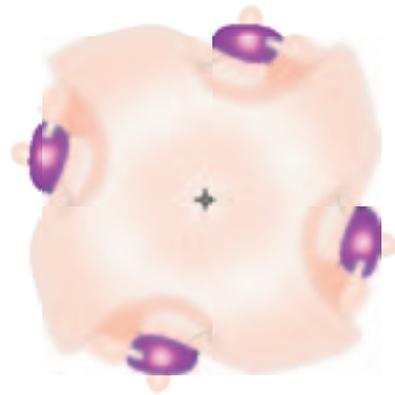
**(b)**

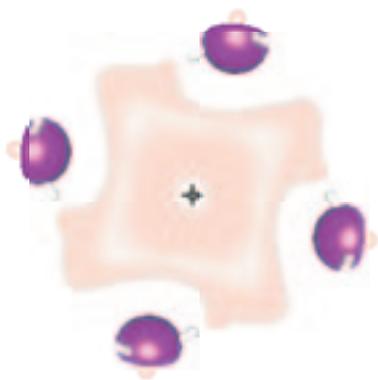
**(c)**

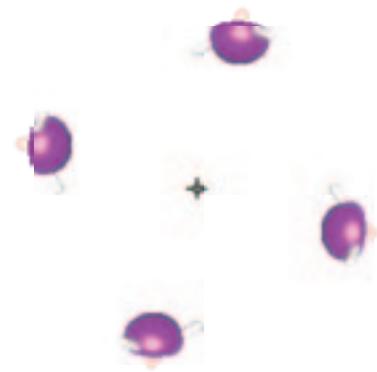
**(d)**

**Figure 1S; Shkrob**

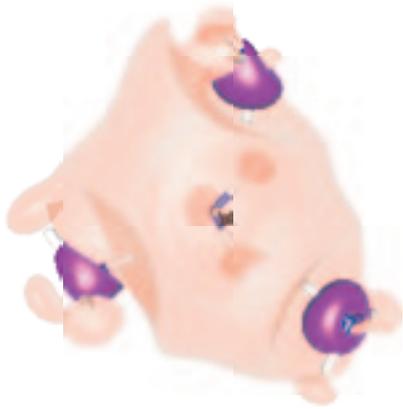

**(a)** **(b)**

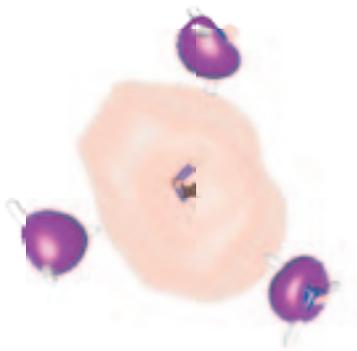 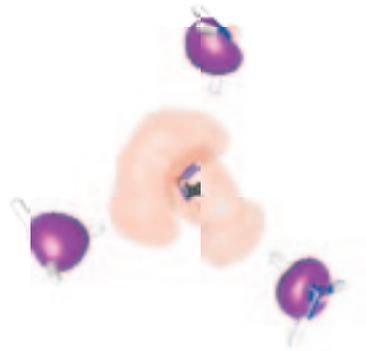

**(c)** **(d)**

**Figure 2S; Shkrob**

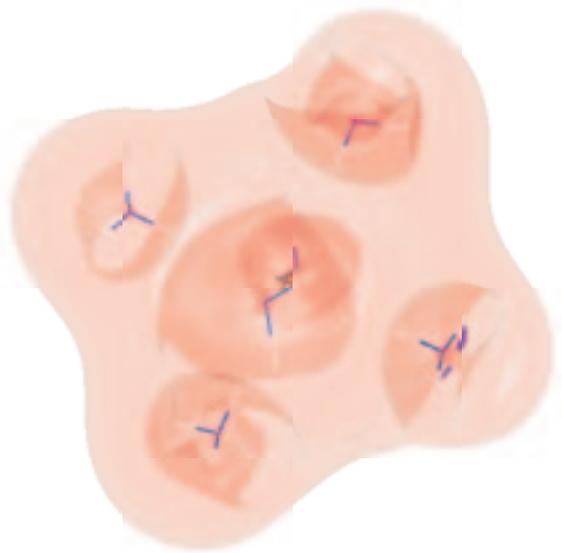
(a)

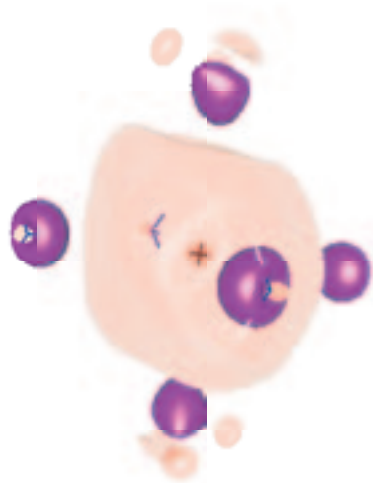
(b)

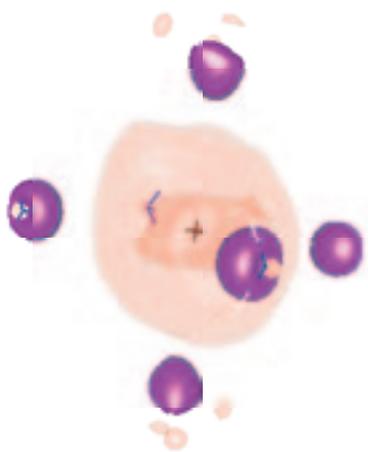
(c)

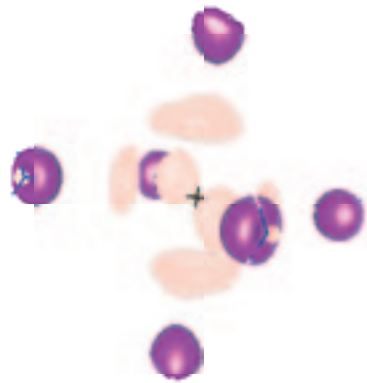
(d)

**Figure 3S; Shkrob**

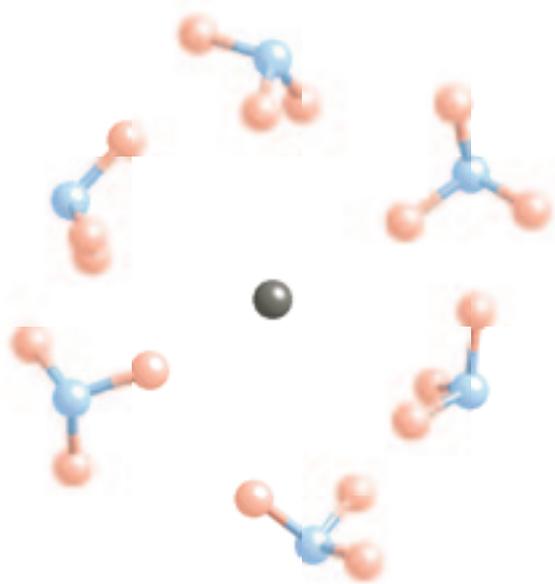

**Figure 4S; Shkrob**

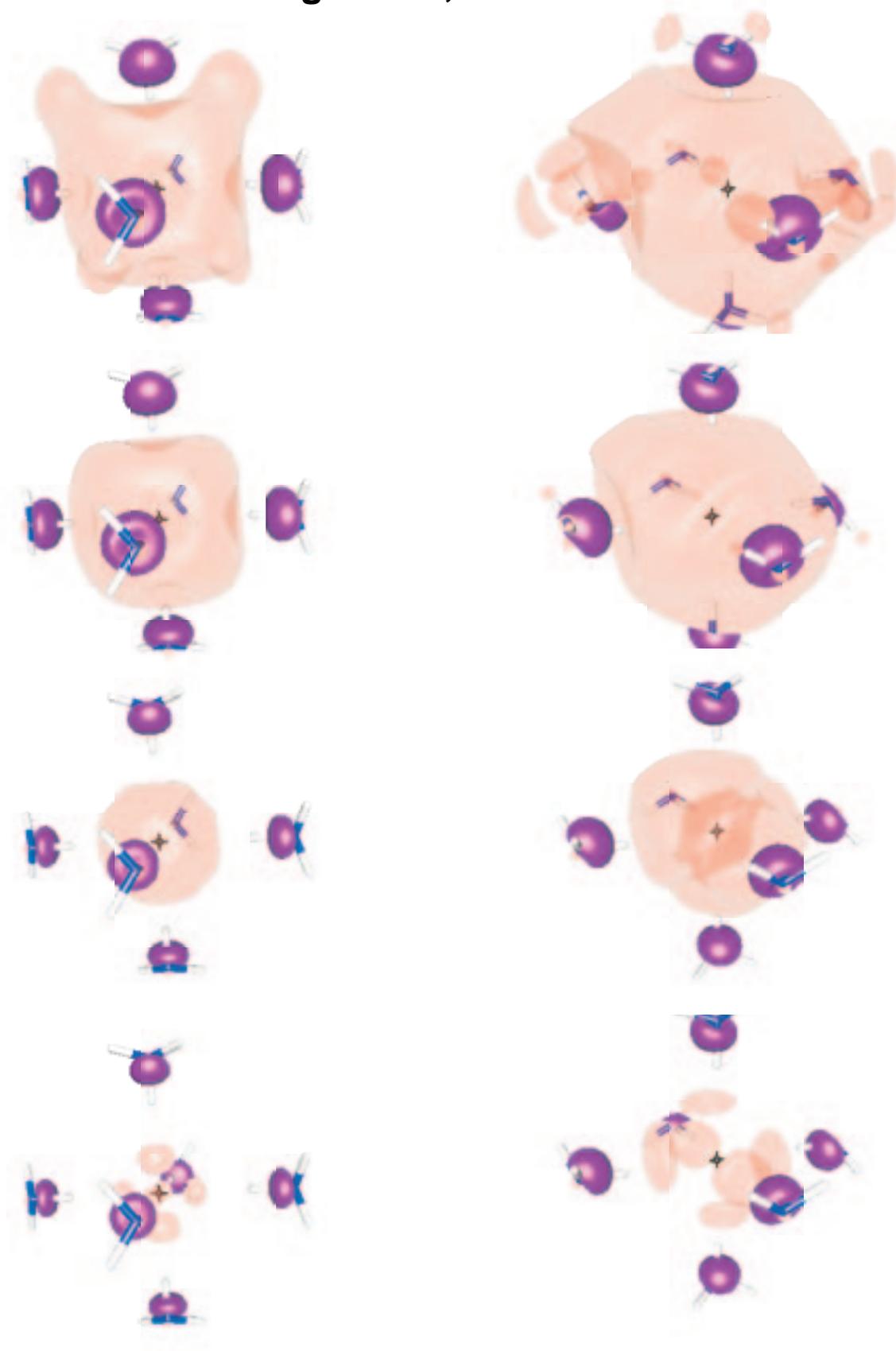

(a) MP2          (b) BLYP